\documentclass[fleqn,usenatbib]{mnras}

\usepackage{newtxtext,newtxmath,hyperref}
\usepackage[T1]{fontenc}

\DeclareRobustCommand{\VAN}[3]{#2}
\let\VANthebibliography\thebibliography
\def\thebibliography{\DeclareRobustCommand{\VAN}[3]{##3}\VANthebibliography}

\usepackage{subcaption}
\usepackage{graphicx}	% Including figure files
\usepackage{amsmath}	% Advanced maths commands

\usepackage{natbib}
\newcommand{\rmu}{rad m$^{-2}$} %Rotation measure units macro
\author[Fine et al.]{
Maxwell  A. Fine,$^{1,2}$\thanks{E-mail: maxwell.fine@mail.utoronto.ca}
Cameron L. Van Eck,$^{3,2}$\thanks{E-mail: cameron.vaneck@anu.edu.au}
and Luke Pratley$^{2}$
\\
$^{1}$Department of Physical and Environmental Sciences, University Of Toronto Scarborough, 1265 Military Trail,Toronto, M1C 1C9, Canada \\
$^{2}$Dunlap Institute for Astronomy and Astrophysics, University of Toronto, 50 St. George Street, Toronto, ON M5S 3H4, Canada \\
$^{3}$Research School of Astronomy \& Astrophysics, The Australian National University, Canberra, ACT 2611, Australia
}

\title[Bandwidth depolarization correction]{Correcting Bandwidth Depolarization by Extreme Faraday Rotation}

\date{Accepted XXX. Received YYY; in original form ZZZ}

\pubyear{2021}

\begin{document}
\label{firstpage}
\pagerange{\pageref{firstpage}--\pageref{lastpage}}
\maketitle

\begin{abstract}

Measurements of the polarization of radio emission are subject to a number of depolarization effects such as bandwidth depolarization, which is caused by the averaging effect of a finite channel bandwidth combined with the frequency-dependent polarization caused by Faraday rotation. There have been very few mathematical treatments of bandwidth depolarization, especially in the context of the rotation measure (RM) synthesis method for analyzing radio polarization data. We have found a simple equation for predicting if bandwidth depolarization is significant for a given observational configuration. We have derived and tested three methods of modifying RM synthesis to correct for bandwidth depolarization. From these tests we have developed a new algorithm that can detect bandwidth-depolarized signals with higher signal-to-noise than conventional RM synthesis and recover the correct source polarization properties (RM and polarized intensity). We have verified that this algorithm works as expected with real data from the LOFAR Two-metre Sky Survey. To make this algorithm available to the community, we have added it as a new tool in the RM-Tools polarization analysis package.

\end{abstract}

\begin{keywords}
polarization -- techniques: polarimetric
\end{keywords}

%%%%%%%%%%%%%%%%% BODY OF PAPER %%%%%%%%%%%%%%%%%%
\section{Introduction}

Observations of polarized emission at radio frequencies enable the study of galactic and extra-galactic magnetic fields via Faraday rotation.
Faraday rotation occurs when linearly polarized\footnote{Throughout the remainder of this work, we implicitly ignore any circularly polarized emission and focus exclusively on linear polarization.} electromagnetic waves pass through a birefringent medium such as magnetized plasma, causing the polarization angles to be altered.
%Faraday rotation can be observed in many astrophysical environments containing magnetized plasma, including the interstellar medium of our Galaxy \citep[e.g.,][]{Brown07} and other galaxies \citep{Beck2015}, star formation regions \citep[e.g.,][]{Tahani2018}, galaxy clusters \citep[e.g.,][]{Bonafede2010}, and potentially the intergalactic medium \citep[e.g.,][]{Vacca2016, Carretti2022}. 
Faraday rotation measurements of background radio sources are typically reported using a quantity called the rotation measure (RM), which gives the strength of the Faraday rotation experienced by emission from each source. RMs contain information on the magnetic fields of the objects along along the line of sight, which can include the Milky Way \citep[e.g.,][]{Jaffe2010}, individual ISM features such as HII regions \citep[e.g.,][]{Costa2015} and molecular clouds \citep[e.g.,][]{Tahani2018}, nearby galaxies \citep{Beck2015}, more distant intervening galaxies \citep[e.g.,][]{Farnes2014}, the intergalactic medium \citep[e.g.,][]{Vacca2016, Carretti2022}, and more. 

The accuracy of these studies depends on the quality and completeness of the RM catalogs they use. Polarization measurements are subject to various depolarization processes, which reduce the polarized intensity in the measured signal. All depolarization processes share the common property of being caused by the superposition of polarized signals with different polarization angles. These processes can be either intrinsic to the emission, such as geometric depolarization caused by different sources of polarized emission along a line of sight having different polarization angles, or instrumental, such as beam depolarization where the finite resolution of a telescope causes adjacent differently-polarized lines of sight to be averaged together. Intrinsic depolarization processes can contain interesting astrophysical information \citep[e.g.,][]{Sokoloff98}, while instrumental depolarization processes must be either accounted for, mitigated, or corrected in order to allow accurate analysis of polarimetric data.

Bandwidth depolarization occurs when a polarized signal with a frequency-dependent polarization angle is averaged over the bandwidth of a frequency channel. Since Faraday rotation by magnetized plasma causes the polarization angle to become frequency dependent, the strength of bandwidth depolarization is closely coupled to the magnitude of the RM. The majority of past observations have not been meaningfully affected by bandwidth depolarization, due to some combination of narrow (in wavelength-squared) channels or low-|RM| targets. As a result, bandwidth depolarization has generally not been considered a major problem requiring correction or accounting for, except for those few cases of observations that were strongly affected by bandwidth depolarization \citep[e.g.,][]{Taylor09}. 

However, bandwidth depolarization is becoming a larger problem for current and future measurements, for three major reasons. First is the increasing use of low (< 300 MHz) observations \citep[e.g.,][]{VanEck2018, Riseley2018} where even narrow-frequency channels span much wider ranges of $\lambda^2$, leading to stronger bandwidth depolarization than higher frequency data. Second, the recent discoveries of very large (> 2000 \rmu) RMs in the Galactic plane \citep{Shanahan2019} and fast radio bursts with extreme RMs \citep{Michilli2018}, which motivate searches for very large RMs that produce correspondingly stronger bandwidth depolarization. Third, the increasing size of radio data sets (from wide-field imaging, wide-band correlators, etc) and the limited signal-to-noise ratio of narrow channels is pushing a desire for broader channels, which come with increased bandwidth depolarization. Bandwidth depolarization is already a significant problem for low-frequency observations such as the LOFAR Two-meter Sky Survey (LoTSS, \citealt{Shimwell2017}), may cause some problems for the upcoming Very Large Array Sky Survey (VLASS, \citealt{Lacy2020}), and is likely to become a problem for future observations with the Square Kilometre Array (SKA). As a result, there is an increasing motivation for developing a method to correct for bandwidth depolarization in those cases where it cannot be mitigated.

\defcitealias{Schnitzeler2015}{SL15}
\defcitealias{Pratley2020}{PJ20}
Since most polarization observations are configured to minimize the effects of bandwidth depolarization, the majority of studies simply neglect it completely. When the bandwidth depolarization is considered in the context of a specific study, few to no details are given about how the depolarization is modelled \citep[e.g.,][]{Condon1998}. In recent years, two papers have worked through theoretical descriptions of bandwidth depolarization: \citet{Schnitzeler2015} and \citet{Pratley2020}. \citet[hereafter SL15]{Schnitzeler2015} derived equations describing bandwidth depolarization and proposed a modified version of the RM synthesis algorithm that accounted for the polarization angle-altering effects of bandwidth depolarization but not the polarized intensity altering effects. \citet[hereafter PJ20]{Pratley2020} also investigated bandwidth depolarization by examining the mathematical similarity to the effects of the primary beam in aperture synthesis, and demonstrated that bandwidth depolarization can not only reduce the measured polarized intensity but will also introduce false Faraday complexity.\footnote{Faraday complexity occurs when polarized emission is distributed over a range of RM values, due to mixing of emission and Faraday rotation along a line of sight.} They proposed a modified method that accounted for the effects of finite bandwidth, recovering the correct RM and polarization angle but not the polarized intensity. To date, there has been no published work that fully corrects for bandwidth depolarization effects and recovers the true astrophysical parameters, especially when RM synthesis methods are used. 

In this paper we derive and test a modified form of RM synthesis that can account for bandwidth depolarization effects, allowing it to correctly identify sources with very strong Faraday rotation and remove the effects of bandwidth depolarization. In Section~\ref{sec:theory} we cover the mathematical formulation of bandwidth depolarization and derive several possible modified algorithms that may correct for it. In Section~\ref{sec:comparison} we test these different algorithms with simulations to verify their performance and determine which is likely to be most effective in practice, then in Section~\ref{sec:algorithm} we describe a method to optimally identify sources with strong bandwidth depolarization and recover their true properties. Section~\ref{sec:tests} describes tests on real data to test the performance of this method;  Section~\ref{sec:conclusions} summarizes our results and conclusions.

\section{Theory}\label{sec:theory}
In the following section, we lay out the mathematical foundations for this paper, beginning with an introduction to Faraday rotation and rotation measure synthesis, deriving a model of bandwidth depolarization, and then proposing modified formulations of RM synthesis that may correct for bandwidth depolarization.

\subsection{Faraday Rotation}
Faraday rotation is characterized by its strength, which is given by the Faraday depth ($\phi$) and defined for a given line of sight as a function of distance ($d$),
\begin{equation}
    \phi(d) = 0.81\; \mathrm{rad \: m}^{-2} \; \int_{0}^{d} \frac{n_e}{\mathrm{cm}^{-3}} \frac{B_\parallel}{\mathrm{\upmu G}} \frac{dr}{\mathrm{pc}}
\end{equation}
where $B_\parallel$ is the magnetic field component parallel to the line of sight (defined positive when the magnetic field is directed towards the observer), $n_e$ is the free electron density, the infinitesimal path length $dr$ is directed along the line of sight, and the integral is computed over the radiation path from the observer to a specified distance $d$.
The effect of the Faraday rotation is to modulate the polarization angle ($\chi$) of polarized emission, by an amount that depends on the Faraday depth at the point of emission:
\begin{equation}
    \chi(\lambda^2,d) = \chi_0 + \lambda^2\, \phi(d)
\end{equation}
where $\chi_0$ is the initial polarized angle and $\lambda$ is the wavelength of the radiation.

In cases where a line of sight contains (or is dominated by) only a single source of polarized emission, and that source emits from a single distance (or over a volume in which the Faraday depth changes minimally), the source is referred to as having an RM equal to the Faraday depth at the distance of the source ($d_\mathrm{src}$),
\begin{equation}
    \mathrm{RM} = \phi(d_\mathrm{src}).
\end{equation}
This is sometimes referred to as the `Faraday-simple' case, in contrast with lines of sight with multiple polarized sources or polarized emission broadly distributed over a range of Faraday depth values, which are called `Faraday-complex'. For the remainder of this work, we focus on the Faraday-simple case except where complexity is explicitly mentioned.

\subsection{Rotation Measure Synthesis}
The mathematical derivation of the rotation measure synthesis technique was first laid out by \citet{Burn1966}, and was further developed into a usable form by \citet{Brentjens2005}.

We can start by defining a complex polarization\footnote{We denote all complex quantities with a tilde symbol.}, $\widetilde{P}$, in terms of the polarized intensity ($P$) and angle, or Stokes $Q$ and $U$, as 
\begin{equation}
    \widetilde{P}= Q + iU =  Pe^{2i\chi }.
\end{equation}
With this notation the Faraday rotated polarized signal from a given distance can be written as:
\begin{equation}
    \widetilde{P}(\lambda^{2},d)= Pe^{2i(\chi_0 + \lambda^2 \phi(d))}= \widetilde{P}_0e^{2i\lambda^2 \phi(d) }
\end{equation}
where the 0 subscript is used to indicate the pre-Faraday rotation quantities. The Faraday rotated polarization can be expressed as the product of two terms: $\widetilde{P}_0$ is the initial polarization, while the $e^{2i\lambda^2 \phi(d)}$ term is the change from Faraday rotation.

The observed polarized signal is a summation of emission produced at all distances, as polarized emission can be produced at any distance along a given line of sight. Since there is a mapping from distance to Faraday depth, this can be rewritten \citep[e.g., Appendix A of ][]{Ordog2019} as a summation over the space of all possible Faraday depths,
%At the observer, for any wavelength the polarization is the superposition of polarized emission at all possible Faraday depths. 
%The term Faraday depth ($\phi$) is used to refer to the space of all possible Faraday rotation values, while RM is used to specify the Faraday rotation actually experienced by polarized emission; with this terminology the observed polarization becomes

\begin{equation}\label{P-lambda}
    \widetilde{P}(\lambda^{2})= \int_{-\infty}^{\infty} \widetilde{P}_0(\phi)e^{2i\lambda^2\phi} d\phi,
\end{equation}
where $\widetilde{P}_0(\phi)$ is all the polarized emission with a given Faraday depth. We note that the variable of integration, $\phi$, describes the space of possible Faraday depths, and is distinct from the Faraday depth function $\phi(d)$.

Equation \ref{P-lambda} has the form of an inverse Fourier transform with $\phi$ as the frequency-like variable and $\lambda^2$ as the time-like one. The polarization as a function of Faraday depth, $\widetilde{P}_0(\phi)$, contains the information about the Faraday rotation properties in the line of sight, and the polarization properties of the emission source.
Rotation measure synthesis is the process of estimating $\widetilde{P}_0(\phi)$ from observed $\widetilde{P}(\lambda^{2})$ through the use of a forward Fourier transform. In practice a number of modifications to the direct Fourier transform are used such as discrete transforms (due to having measurements in discrete channels), and channels weights, $w_i$ (to accommodate different sensitivity levels across channels).
%; and a shift to a reference wavelength, $\lambda^2_0$ (to improve the phase/polarization angle behaviour of the resulting function). 
The resulting equation for the observed Faraday dispersion function (FDF), which is the reconstruction of $\widetilde{P}_0(\phi)$ given available data, has the form
\begin{equation}\label{eq:RMsynthesis}
    \widetilde{F}(\phi) = \left(\frac{1}{\sum_i w_i}\right) \sum_i w_i \tilde{P}_i e^{-2i\lambda_i^2\phi}
\end{equation}
where all quantities subscripted with $i$ are the channelized-version of the appropriate parameter. 
%Unless otherwise specified, for the remainder of this paper we will drop the $\lambda^2_0$ as it can be propagated through all relevant equations without introducing significant changes. 
Due to the limited sampling in the $\lambda^2$ domain, the FDF is affected by a transfer function, the rotation measure spread function (RMSF), given by 
\begin{equation}
\mathrm{RMSF} = \left(\frac{1}{\sum_i w_i}\right) \sum_i w_i e^{-2i(\lambda_i^2-\lambda^2_0)\phi}.
\end{equation}

In the derivations that follow, we limit our scope to the Faraday-simple case with a single source of polarization with a well-defined RM. This allows us to make the distinction between the Faraday depth of the source (the RM) and the variable $\phi$ which enables functions like the FDF to span the space of possible Faraday depth values.

%In the derivations that follow, we maintain a careful distinction between RM and $\phi$: the RM is a fixed property of the source, the true amount of Faraday rotation that the observed polarized emission experiences, while the Faraday depth $\phi$ is, specifically in the context of FDFs and related quantities, a free parameter that spans the space of possible RMs. Another way of expressing this distinction is that in the following equations the RM is used when the true properties of the source are involved, while $\phi$ is used for calculations made with possible RM values (which may or may not be the true RM of the source).

\subsection{Bandwidth Depolarization}\label{sec:depol}
By necessity, the frequency channels in a radio observation have finite non-zero bandwidth. Since the polarized signal has a frequency dependence, as a result of Faraday rotation, this requires that the reported channel values are some form of average over the bandwidth of each channel.

\citetalias{Schnitzeler2015} described the effects of bandwidth-averaging using two models of the channel bandpass. One model was a simple top-hat function in the $\lambda^2$ domain (equal sensitivity for all $\lambda^2$ values inside the channel, zero otherwise), which is well-behaved and produces an analytic solution when the effects of Faraday rotation are integrated over such a channel; this model was also used by \citet{Brentjens2005} and \citetalias{Pratley2020}. Their second model, which they use throughout their paper, is a top-hat function in the frequency domain. This is a closer representation of how channels are formed in radio telescopes, so we have also adopted this model for the remainder of this paper. \citetalias{Pratley2020} developed a general formalism with which any bandpass model can be similarly developed.

For a channel with a top-hat bandpass, with a center frequency $\nu_i$ and full width $\Delta\nu_i$, the measured polarization (for line of sight with a single source of polarization) in a given channel, assuming no intrinsic frequency-dependence other than Faraday rotation (i.e., ignoring spectral index), is
\begin{align}
    \widetilde{P}(\nu) = & \frac{1}{\Delta\nu_i} \int_{\nu_i-\Delta\nu_i/2}^{\nu_i+\Delta\nu_i/2} \widetilde{P}_0 e^{2i \lambda^2 \,\mathrm{RM}} d\nu \label{eq:channel_pol} \\
    = & \widetilde{P}_0  \int_{\nu_i-\Delta\nu_i/2}^{\nu_i+\Delta\nu_i/2} e^{2i (c/\nu)^2\,\mathrm{RM}} \frac{d\nu}{\Delta\nu_i} \nonumber \\
    = & \widetilde{P}_0 \, R_i(\nu,\mathrm{RM}), \label{eq:R_i}
\end{align} 
where this integral is the channel-averaged Faraday rotation operator, which we label as $R_i(\nu,\mathrm{RM})$. The indefinite form of this integral has the solution \citepalias[adapted from][Equation 13]{Schnitzeler2015} 

\begin{multline}
    \int e^{2i(c/\nu)^2\,\mathrm{RM}} d\nu = \nu e^{2i(c/\nu)^2\,\mathrm{RM}} + \\ 
    c \sqrt{|\mathrm{RM}|\pi}\left[ \mathrm{sign}(\mathrm{RM})-i \right]\mathrm{erf}\left(\sqrt\phi\frac{c}{\nu}(\mathrm{sign}(\mathrm{RM})-i )\right).    
\end{multline}

The definite integral does not simplify. In the limit of zero bandwidth it reduces down to only the Faraday rotation, as expected. With finite bandwidth, there are shifts in the polarization angle (reflecting how the Faraday rotation is asymmetric in the frequency domain) and reductions in the polarized intensity (reflecting the ``vector-averaging'' nature of combining polarized signals with different polarization angles). This last effect is bandwidth depolarization, and causes reduced signal-to-noise ratio in cases where the bandwidth and/or Faraday depth are large.

\begin{figure}
	\includegraphics[width=\columnwidth]{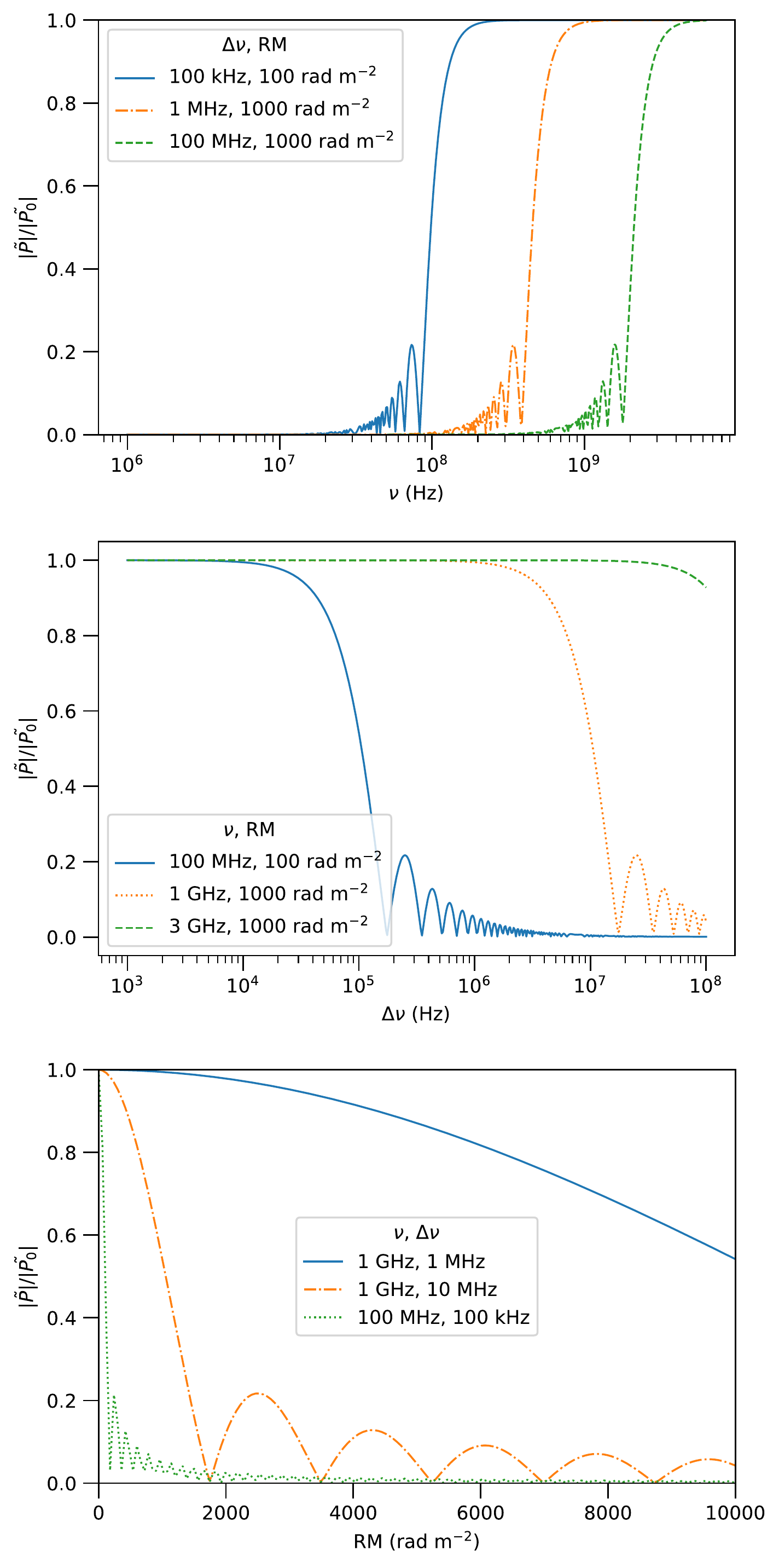}
    \caption{Examples of how strong bandwidth depolarization is (expressed as a ratio of measured to true intensity polarized intensity) as a function of frequency (top panel), channel bandwidth (center panel), and RM (bottom panel). For each panel, the various lines show values selected to show representative values for large disk (1000~\rmu) and large halo/typical disk (100~\rmu) RMs, and frequency and bandwidth configurations typical of different kinds of polarization observations.
      }
    \label{fig:rotation_operator}
\end{figure}

Using Eq.~\ref{eq:channel_pol} it is possible to predict the strength of the bandwidth depolarization, expressed as the fraction of the true polarization that is measured ($|\widetilde{P}|/|\widetilde{P}_0|$), for any combination of channel frequency, bandwidth, and RM. Fig.~\ref{fig:rotation_operator} shows some examples of the strength of bandwidth depolarization for various combinations of these parameters. Three frequency/bandwidth combinations are shown: 1 GHz with 1 MHz channels, which is fairly typical of L-band observations with fine channelization (e.g., with the Australian Square Kilometre Array Pathfinder; \citealt{Hotan2021}); 1 GHz with 10 MHz channels, representing an L-band observation with coarser channels (e.g., with the Very Large Array; \citealt{Perley2011}), and 100 MHz with 100 kHz channels, representing values from a low-frequency instrument (e.g., the Low Frequency Array or Murchison Widefield Array; \citealt{vanHaarlem2013, Tingay2013}). Two reference values of RM are used: 100~\rmu\ to represent values are typical in the Galactic disk and large but possible in the halo, and 1000~\rmu\ to represent large but possible values in the Galactic disk. If we consider bandwidth depolarization to be significant when there is a greater than 10\% loss of polarized signal (a measured-to-true polarization ratio of 0.9 or less, in Fig.~\ref{fig:rotation_operator}), then some of the observational configurations will have populations of sources that are significantly bandwidth depolarized.

The lack of a closed-form solution for single-channel bandwidth depolarization makes it challenging to find a simple calculation that can be used to estimate the strength of bandwidth depolarization on any given calculation. We have numerically explored a wide range of the parameter space, and found some approximate scaling relationships that appear reliable in the weak-depolarization regime ($|\widetilde{P}|/|\widetilde{P}_0| \gtrsim 0.5$). These relationships are most straightforwardly expressed as the RM at which the depolarization reaches a given threshold, 
\begin{equation}\label{eq:C}
    \mathrm{RM} = C \, \left( \frac{\nu_i}{\mathrm{GHz}} \right)^3 \left( \frac{\Delta\nu_i}{\mathrm{MHz}} \right)^{-1}
\end{equation}
where $C$ is a constant determined by the selected threshold in depolarization.\footnote{This  dependence on frequency cubed and the inverse of bandwidth can be independently derived from equation 63 of \citet{Brentjens2005}, which expresses the same relation in terms of the channel's width in the $\lambda^2$ domain. Their value for the constant $C$ is 9636 \rmu\ for 50\% depolarization, which is approximately consistent with our value of 10545 \rmu\ given the different channel shape model used.}
We report some typical values for $C$ in Table~\ref{tab:C}; these values appear to be reliable within 1\% across the regions of the parameter space where the fractional bandwidth ($\Delta\nu / \nu$) is less than 0.1. We propose that this equation can be used as a convenient estimator for determining if bandwidth depolarization is significant for a dataset: an observer can input the frequency and bandwidth of their lowest-frequency channel, chose a depolarization threshold that is acceptable to them, and determine if the corresponding RM for that threshold is larger than the Faraday depth search space they wish to explore.

\begin{table}
\centering
 \caption{Values of RM limit constant $C$ for different depolarization thresholds}
 \label{tab:C}
 \begin{tabular}{lc}
  \hline
  $|\widetilde{P}|/|\widetilde{P}_0|$ & $C$ (\rmu)\\
  \hline
    0.99 & 1360 \\
    0.95 & 3070 \\
    0.90 & 4380 \\
    0.80 & 6290 \\
    0.50 & 10550\\
  \hline
 \end{tabular}
\end{table}

\subsection{Deriving corrections for bandwidth depolarization}\label{sec:derivations}
The goal of this work was two-fold: to develop a method that could identify polarized sources strongly affected by bandwidth depolarization with the greatest possible signal-to-noise ratio (S:N), and to correctly extract the true values (i.e., before bandwidth depolarization) of the polarized intensity and RM.

Below we derive equations for the measured polarized intensity and noise for conventional RM synthesis as well as 3 modified transformations that we predicted to have potentially useful properties.

\subsubsection{Conventional RM synthesis}\label{sec:conventional_theory}
Combining equations \ref{eq:channel_pol} and \ref{eq:RMsynthesis}, the FDF for an ideal Faraday-simple, bandwidth-depolarized source takes the form
\begin{equation}
        \widetilde{F}_\mathrm{synth}(\phi) = \left(\frac{1}{\sum_i w_i}\right) \sum_i w_i \left(\widetilde{P}_0 \, R_i(\nu,\mathrm{RM})\right) \, e^{-2i\lambda_i^2\phi}.
\end{equation}
In the zero-bandwidth limit $R_i$ reduces to the basic Faraday rotation operator, $e^{2i\lambda_i^2\mathrm{RM}}$, and it can be seen that when $\phi = \mathrm{RM}$ all channels will be in-phase and will add constructively, producing the maximum amplitude signal of $\widetilde{P}_0$ and the corresponding peak in the FDF as normally expected.

However, in the bandwidth-depolarized case there is no analytic equation for the resulting polarized intensity, as the channel-averaged polarization angles will not be aligned with the expected channel-center angles as previously described, so that the corresponding phase-cancellation will not occur. However, an upper limit can be found by ignoring these phase effects and considering only the loss of polarized signal within the channels. We define the sensitivity, $S(\mathrm{RM})$ as the ratio of measured polarized intensity to true polarized intensity:
\begin{equation}\label{eq:Q}
    S_\mathrm{synth}(\mathrm{RM}) = \frac{|\widetilde{F}(RM)|}{|\widetilde{P}_0|} \le \left(\frac{1}{\sum_i w_i}\right) \sum_i w_i  |R_i(\nu,\mathrm{RM})|.
\end{equation}
We note that this sensitivity upper-bound assumes that a peak occurs at the Faraday depth corresponding to the RM of the source. Figure 1 of \citetalias{Schnitzeler2015} and figure 10 of \citetalias{Pratley2020} demonstrate that this is generally not the case when bandwidth depolarization is very strong, and that multiple adjacent peaks may instead be produced.

The theoretical noise level can also be predicted. We assume that both Stokes $Q$ and $U$ have Gaussian-distributed noise with same noise standard deviation, $\sigma_{QU,i}$, for each channel, but not necessarily the same standard deviation for all channels, and that the noise in each channel is independent of other channels. With the first assumption, the noise distribution is rotationally invariant, and thus unchanged by any polarization angle changes and is only affected by scaling. The variance in the FDF can be derived as the sum of variances of each channel, multiplied by the appropriate scaling factors:
\begin{equation}
    \sigma_\mathrm{synth}^2(\phi) = \left(\frac{1}{\sum_i w_i}\right)^2 \sum_i w_i^2 \, \sigma_{QU,i}^2.
\end{equation}
In this case the noise is independent of Faraday depth, but this will not be true for the following cases.

\subsubsection{Direct inverse transform}
The first apparent solution to the problems bandwidth depolarization causes in the conventional RM synthesis method is to replace the Fourier/Faraday-derotation kernel with an operator that accurately removes the channel-averaged Faraday rotation effects, including the bandwidth depolarization. This would be the inverse of $R_i(\nu,\phi)$, leading to a modified FDF:
\begin{equation}
    \widetilde{F}_\mathrm{inverse}(\phi) = \left(\frac{1}{\sum_i w_i}\right) \sum_i w_i \left(\widetilde{P}_0 \, R_i(\nu,\mathrm{RM})\right) \, R_i(\nu,\mathrm{\phi})^{-1}\\    
\end{equation}
For the Faraday depth equal to the source RM, the rotation and inverse operators cancel out and the original polarization state $\widetilde{P}_0$ is recovered; in the noise-free case this transform has the property $S_\mathrm{inverse}(\mathrm{RM}) = 1$. The theoretical noise level is no longer independent of Faraday depth:
\begin{equation}
    \sigma_\mathrm{inverse}^2(\phi) = \left(\frac{1}{\sum_i w_i}\right)^2 \sum_i w_i^2 \, \sigma_{QU,i}^2 |R_i(\nu,\mathrm{\phi})|^{-2}.
\end{equation}
The magnitude of the rotation operator, $|R_i(\nu,\mathrm{\phi})|$, is strictly equal to or less than 1, so the noise is significantly amplified if there are channels that are strongly depolarized. In the hypothetical worst case where there is complete depolarization, the FDF and the noise both suffer from division by zero and become undefined; in the more realistic case of arbitrarily strong (but not complete) depolarization the noise can be amplified to an arbitrarily high degree. This introduces problems with identifying the true RM of a source (or identifying if the source even has significant polarization), because the highest amplitude peak in the FDF can easily be a noise peak in a part of the FDF where the noise is strongly amplified. The inverse transform is theoretically effective at recovering the correct polarized intensity, but only if the true RM value is known {\it a priori.}

One additional concern, which is applicable to all three modified transforms, is that these are no longer Fourier transforms and the corresponding Fourier properties that are often used in RM synthesis no longer apply, so some care must be taken in applying conventional properties of RM synthesis. One such property that complicates matters is that the RMSF is no longer well defined: emission with different RMs will not produce the same shape in the FDF, as is the case for RM synthesis, but will instead have a different convolved transfer function as a function of both RM and $\phi$. This is derived in more detail in a later section.

\subsubsection{SL15 transform}
\citetalias{Schnitzeler2015} proposed a modified transform of a similar type, but instead of inverting the full effects of the channel-averaged Faraday rotation only the phase/polarization angle effects were reversed. In terms of the formalism used in this paper, their operator was $|R_i(\nu,\mathrm{\phi})|/R_i(\nu,\mathrm{\phi})$, which has an amplitude of 1 and the same phase as $R_i(\nu,\mathrm{\phi})$. The transform is then defined as 
\begin{equation}
    \widetilde{F}_\mathrm{SL15}(\phi) = \left(\frac{1}{\sum_i w_i}\right) \sum_i w_i \left(\widetilde{P}_0 \, R_i(\nu,\mathrm{RM})\right) \, |R_i(\nu,\mathrm{\phi})| \, R_i(\nu,\mathrm{\phi})^{-1},    
\end{equation}
the sensitivity as
\begin{equation}
    S_\mathrm{SL15}(\mathrm{RM}) = \left(\frac{1}{\sum_i w_i}\right) \sum_i w_i  |R_i(\nu,\mathrm{RM})|,
\end{equation}
and the noise as
\begin{equation}
    \sigma_\mathrm{SL15}^2(\phi) = \left(\frac{1}{\sum_i w_i}\right)^2 \sum_i w_i^2 \, \sigma_{QU,i}^2.
\end{equation}
This transform behaves very similarly to the conventional RM synthesis transform, with similar noise behaviour, and the sensitivity equal to the upper limit of conventional RM synthesis (as that upper limit was estimated by assuming the phases were all aligned, whereas here they are explicitly made to be aligned). \citetalias{Schnitzeler2015} demonstrated that this transform produces higher polarized intensity values (and correspondingly better S:N) at large RMs compared to the conventional transform, although these polarized intensities are still not the true values.

\subsubsection{Adjoint transform}
The third transform we considered was an adjoint transform, which was used by \citetalias{Pratley2020}, which is defined as the operator $R_i^*(\nu,\phi)$, where the asterisk indicates a Hermitian conjugate. The adjoint transform is commonly seen in the context of optimal cosmological map making and it is used in the process of least squares fitting. In particular, for ill-posed problems the adjoint is sometimes the only choice (e.g., a masking operation has no inverse) and it can avoid boosting the noise when the inverse is ill-conditioned.  The transform has the form
\begin{equation}\label{eq:adjointFDF}
    \widetilde{F}_\mathrm{adj}(\phi) = \left(\frac{1}{\sum_i w_i}\right) \sum_i w_i \left(\widetilde{P}_0 \, R_i(\nu,\mathrm{RM})\right) \, R_i^*(\nu,\mathrm{\phi}).    
\end{equation}
This operator has the effect of reversing the phase shift of the Faraday rotation, but not correcting the loss of amplitude from the bandwidth depolarization. Instead, the amplitude is further reduced by a factor of $|R_i^*(\nu,\mathrm{\phi})|$, which has the net effect of reducing the contribution from (down-weighting) channels where the depolarization is predicted to be strong and the resulting signal would be weak. The resulting sensitivity is
\begin{equation}\label{eq:adjointsensitivity}
    S_\mathrm{adj}(\mathrm{RM}) = \left(\frac{1}{\sum_i w_i}\right) \sum_i w_i  |R_i(\nu,\mathrm{RM})|^2,
\end{equation}
and the noise is
\begin{equation}\label{eq:adjointnoise}
    \sigma_\mathrm{adj}^2(\phi) = \left(\frac{1}{\sum_i w_i}\right)^2 \sum_i w_i^2 \, \sigma_{QU,i}^2 |R_i(\nu,\mathrm{RM})|^2.
\end{equation}
This transform has the opposite behaviour as the direct inverse transform, in that the noise decreases as the bandwidth depolarization becomes stronger, which this makes it much better behaved in cases of very strong depolarization.

\section{Comparison of transform equations}\label{sec:comparison}
To empirically test the behaviour of the different transforms, we constructed a series of simulations of bandwidth depolarized sources and observed the properties of the resulting FDFs for each of the four transforms. For this section, each simulation was of a single Faraday-simple source, observed with 488 channels of 96 kHz bandwidth each from approximately 120 to 170 MHz. These frequencies were selected to match the LoTSS survey, which is expected to have strong bandwidth depolarization in the Galactic disk. For this frequency range and channel bandwidth, bandwidth depolarization is expected to become significant for RMs of a few hundred \rmu, while the RMSF width is approximately 1 \rmu. For each channel, Gaussian random noise was added separately to Stokes $Q$ and $U$; the noise variance was the same for all channels, but was adjusted between simulation runs to control the S:N. Stokes $I$ spectra were not included in the simulation. For all the results below, we also ran simulations with other frequency configurations (e.g., VLASS-like, and the expected frequency configuration of the Polarization Sky Survey of the Universe's Magnetism, POSSUM; \citealt{Gaensler2010}) to check for any behaviour differences; all tests produced the same results (subject to the scaling relationships described in Sec.~\ref{sec:depol}), so we consider the conclusions valid for all observation configurations consisting of a single band of equal-sized channels. Observations consisting of multiple separated bands and/or with variable channel-widths have not been explored, but on theoretical grounds we do not expect them to behave significantly differently.

The transforms were implemented by modifying the RM synthesis package \textsc{RM-Tools} \citep{Purcell2020}. The \textsc{RM-Tools} \texttt{rmsynth1d} tool implements conventional RM synthesis, followed by a peak-characterization algorithm that identifies the highest amplitude peak in the FDF, fits a quadratic function to that peak, extracts the RM and polarized intensity of the peak from that fit, then estimates other parameters such as the polarization angle and uncertainties on all quantities.

\subsection{Comparing Faraday Dispersion Functions}
% 4 simlated FDS
% make this plot larger, full page?
% maybe put in 2 next to each other

\begin{figure*}
	\includegraphics[width=\textwidth,height=0.9\textheight,keepaspectratio]{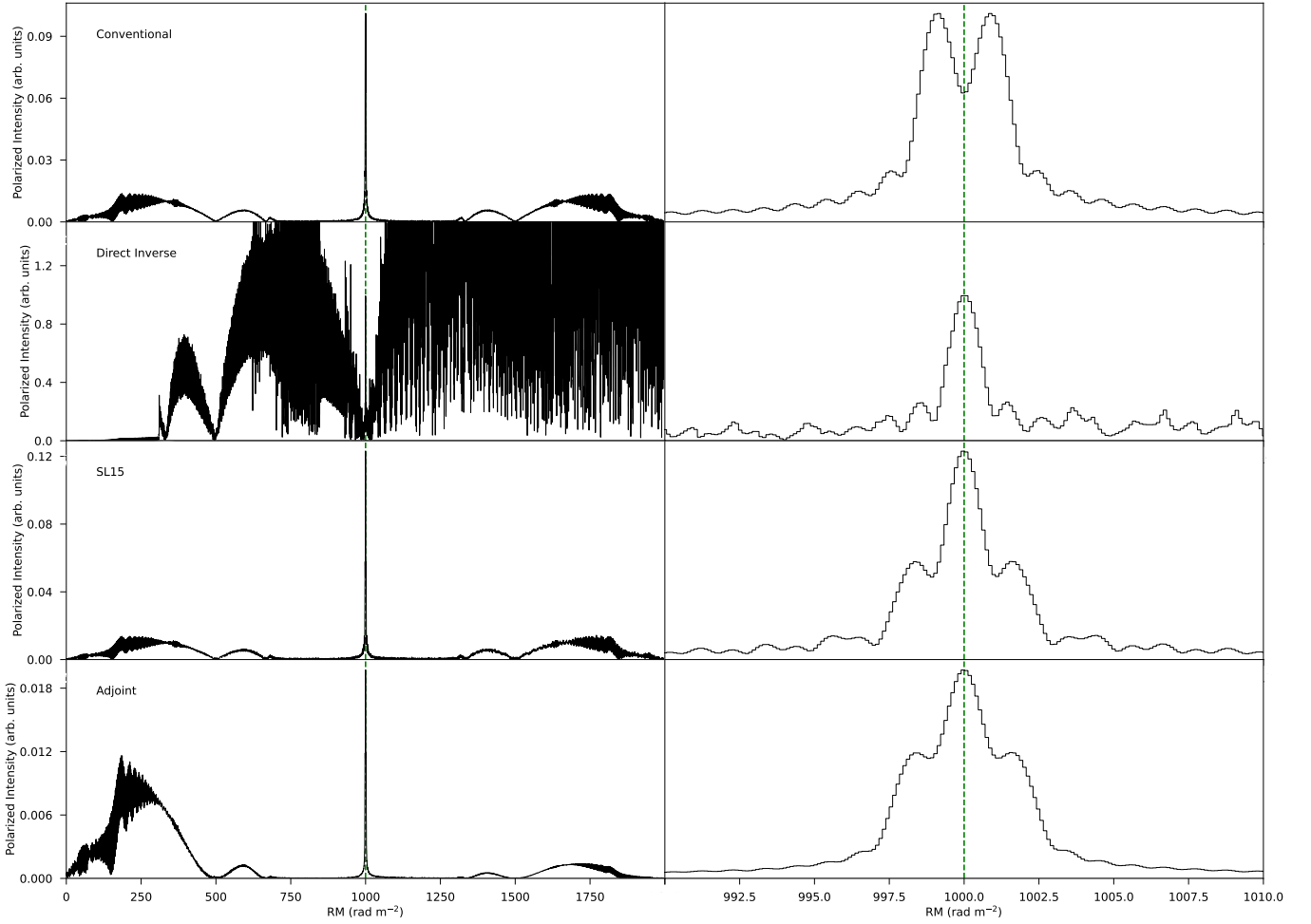}
    \caption{Comparison of the four Faraday Dispersion Functions (FDFs) produced by the four algorithms over the same simulated data set (RM = 1000~\rmu, polarized intensity = 1); the left panels show the FDFs from 0 to 2000~\rmu, the right panels show $\pm$10~\rmu\ around the true peak. For each, the true RM (the RM of the simulated source) is marked by a green dashed line.
      }
    \label{fig:Faraday Disperson Functions}
\end{figure*}

We show an example of the Faraday depth function produced by each of the four transforms described above, for a strongly bandwidth depolarized source, in Fig.~\ref{fig:Faraday Disperson Functions}. The simulated source had an RM of 30000 \rmu\ (much larger than typically observed, but chosen to make the differences between transforms dramatically apparent), polarized intensity of 1, and noise with standard deviation of 0.03 (in each channel and Stokes parameter). The four FDFs produced by the different algorithms have significant differences. The conventional algorithm has a twin peak bracketing the true Faraday depth value, causing the peak detection algorithm to identify only one of the peaks and produce a corresponding incorrect measured RM. The amplitude of the two peaks is substantially reduced by bandwidth depolarization, from a true value of unity to 0.08. The direct inverse transform does produce a peak with the correct location and amplitude but the spectrum is dominated by noise peaks, especially at large RMs. This causes the peak-characterizing algorithm to focus on a noise peak instead of the correct peak. The \citetalias{Schnitzeler2015} transform has a single sharp peak located at the true Faraday depth, but the amplitude was still strongly affected by the bandwidth depolarization. The adjoint algorithm performs similarly to the \citetalias{Schnitzeler2015} algorithm in that it has a single sharp peak located at the true Faraday Depth, albeit with even more reduced polarized intensity, and also has the unexpected property of having somewhat smoother sidelobe structure.

The behaviour of the FDFs does follow the expectations described in the derivations of Sec~\ref{sec:derivations}. However, it is clear that the modified transforms do not interact well with the peak-characterization algorithm used by \textsc{RM-Tools}, which simply looks for the maximum polarized intensity in the FDF. When analyzing sources without an {\it a priori} known RM, an alternative method of identifying the most significant peak in the FDF must be developed.

\subsection{Comparing Sensitivity}
\begin{figure}
	\includegraphics[width=\columnwidth]{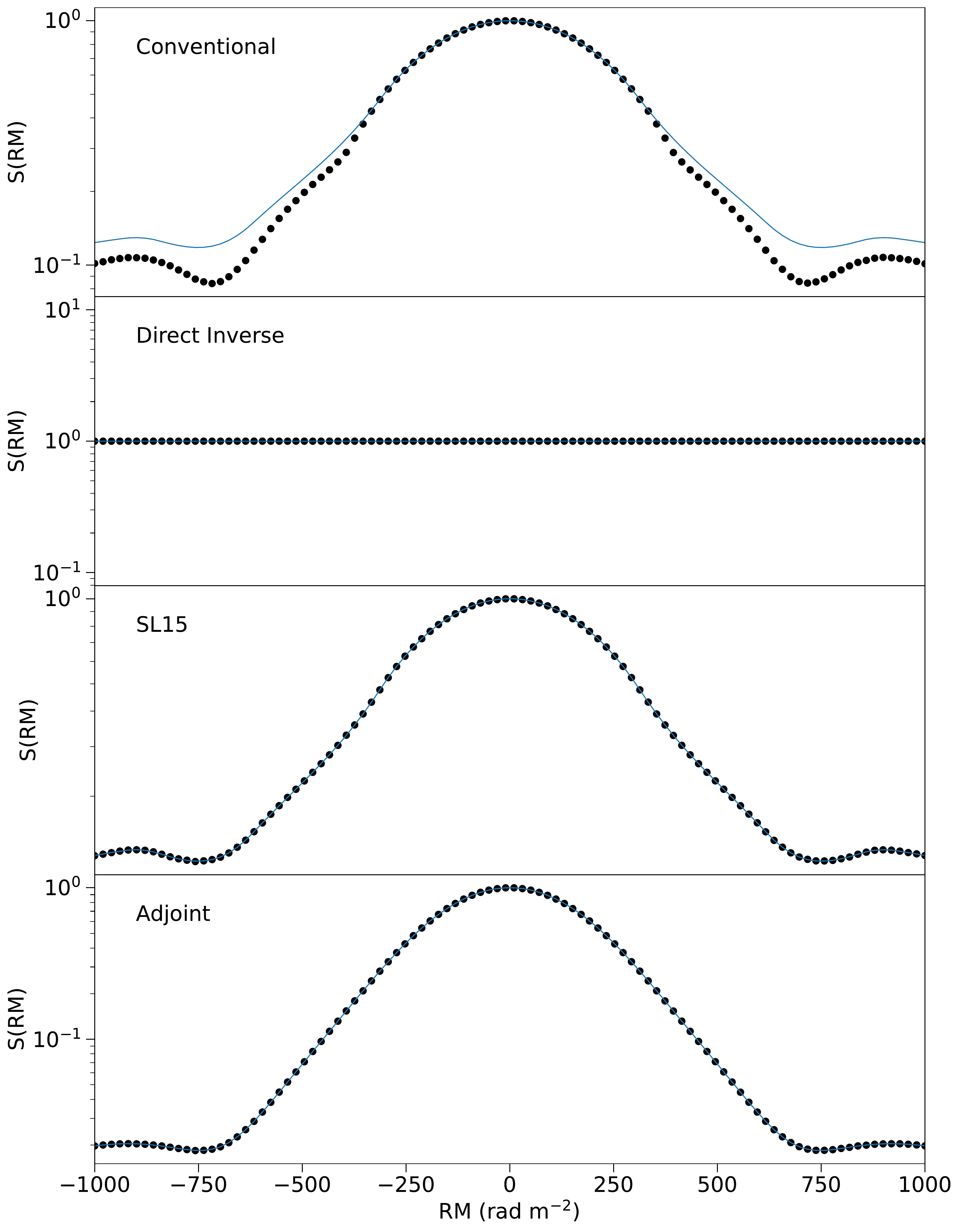}
    \caption{Comparison of the sensitivity ($S(\mathrm{RM})$, ratio of measured polarized intensity to true polarized intensity) for the four transforms, showing both the theoretical sensitivity (blue lines) and empirically observed sensitivity (black points). The theoretical sensitivity for conventional RM synthesis is an upper limit; the true values are observed to be lower for large |RM| values. In all cases, the empirical results agree with the predictions.
      }
    \label{fig:sensitivity}
\end{figure}

We verified the accuracy of the theoretical sensitivity equations for each transform by constructing a series of noise-free simulations, varying the true RM and measuring the peak polarized intensity in the FDF. In the case of the direct inverse transform, which suffers from false peaks at large RMs even in the absence of noise, we limited the peak-finding algorithm to only search in proximity to the true RM. The results of these tests are shown in Figure~\ref{fig:sensitivity}.

The simulation results generally confirm the predictions of the theoretical sensitivity equation. The conventional transform can only predict an upper limit to the sensitivity (due to residual polarization angle mis-alignments as described in Sec.~\ref{sec:conventional_theory}), and the simulations confirm that at large RMs the recovered polarized intensity is lower than this upper limit, by potentially as much as 40\%. This confirms that it is not possible to accurately correct values after conventional RM synthesis with an analytic correction; a numerical correction, effectively reproducing the simulation we performed, would be required. The direct inverse transform produces the expected sensitivity, which has no dependence on Faraday depth, although this requires that the correct peak be identified even when it is not the strongest peak (as the direct inverse transform creates many strong false peaks). The \citetalias{Schnitzeler2015} and adjoint transforms show simulated behaviour that follows the theoretical predictions very closely; even at very large RMs the ratio of simulated to theoretical sensitivity remains within 2\% of unity.

With the exception of the conventional transform, the theoretical sensitivity curves are able to correctly predict the amount of depolarisation and thus can be used to correct for bandwidth depolarization. By dividing the obtained FDF by the sensitivity curve, a corrected sensitivity spectrum can be computed. The corrected sensitive spectrum has the polarized intensity (PI) of the emission before the effects of depolarization but also suffers from increased noise, particularly at large RMs. Therefore, this corrected sensitivity spectrum is probably not suitable for finding peaks.

\subsection{Comparing Noise}
\begin{figure}
	\includegraphics[width=\columnwidth]{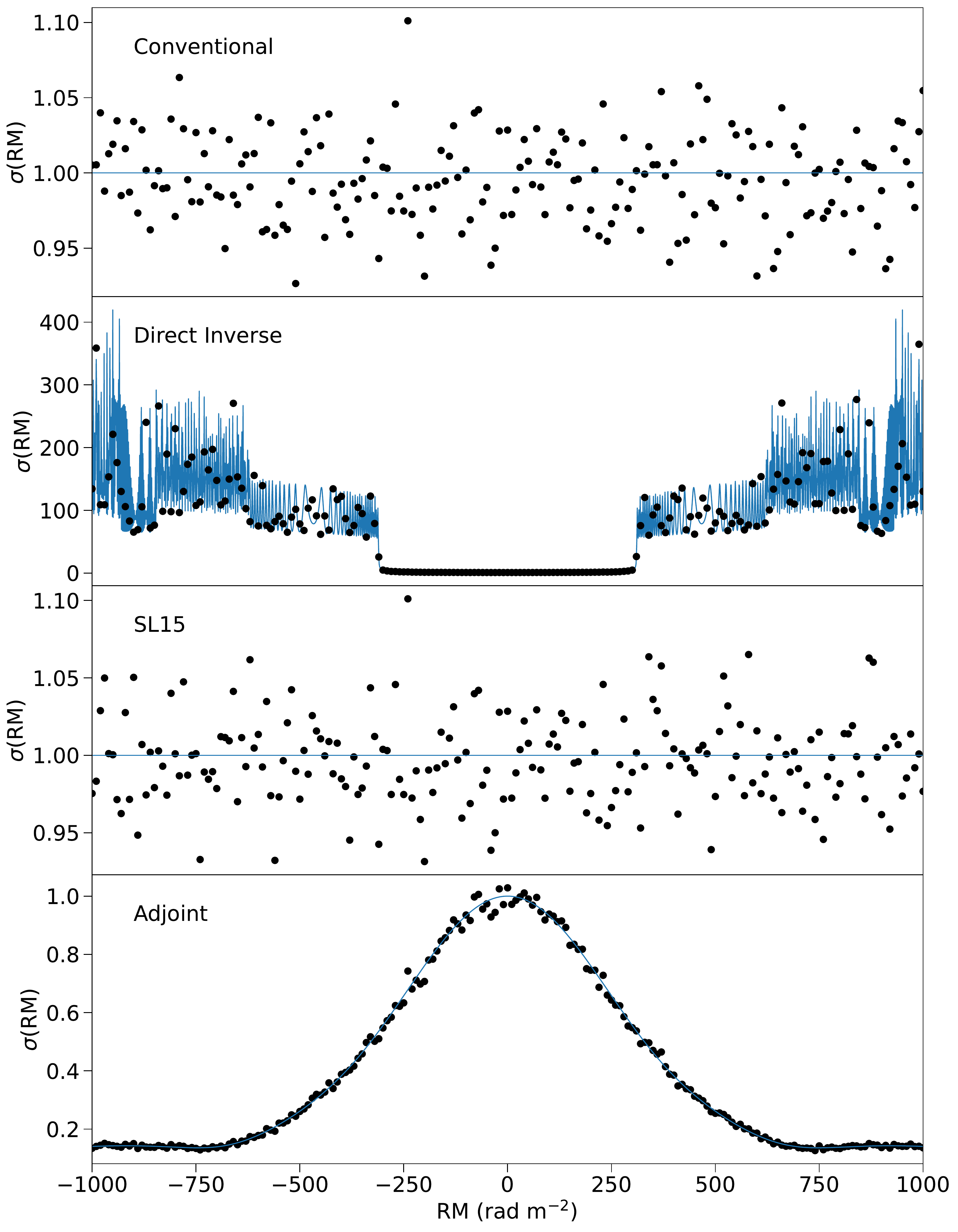}
    \caption{Comparison of the noise level as a function of Faraday depth for the four transforms, showing both the theoretical noise (blue lines) and empirically observed noise (black points). The scatter in empirical points reflects the statistical noise from the limited number of simulation iterations. The empirical results confirm the theoretical predictions for all transforms across the full Faraday depth range sampled.
      }
    \label{fig:noise}
\end{figure}

Similarly to the sensitivity analysis, we verified the noise behaviour of the four transforms by constructing simulations and analyzing the resulting FDFs. These simulations used the same channel configuration as the previous simulations but contained no signal, only noise drawn from a Gaussian distribution for each channel and Stokes parameter. The noise amplitude was chosen so as to make the noise in the conventional FDF equal to 1. To generate estimates of the noise as a function of Faraday depth, we generated 3000 simulated spectra and computed the corresponding FDFs for all four transforms. For each sampled Faraday depth and transform we computed the standard deviation over the ensemble of simulated FDFs (both the real and imaginary components were included when computing the standard deviation). The resulting empirical noise curves are shown in Fig.~\ref{fig:noise}. The scatter seen in the conventional and \citetalias{Schnitzeler2015} transforms is due to the finite number of simulations used compute the empirical values (the amplitude of the scatter was verified to scale as expected with respect to the number of simulations). Otherwise the simulations closely follow the theoretical predictions, confirming that the theoretical noise curves can be used to predict the noise as a function of Faraday depth.

\subsection{Comparing signal to noise ratio}
\begin{figure}
	\includegraphics[width=\columnwidth]{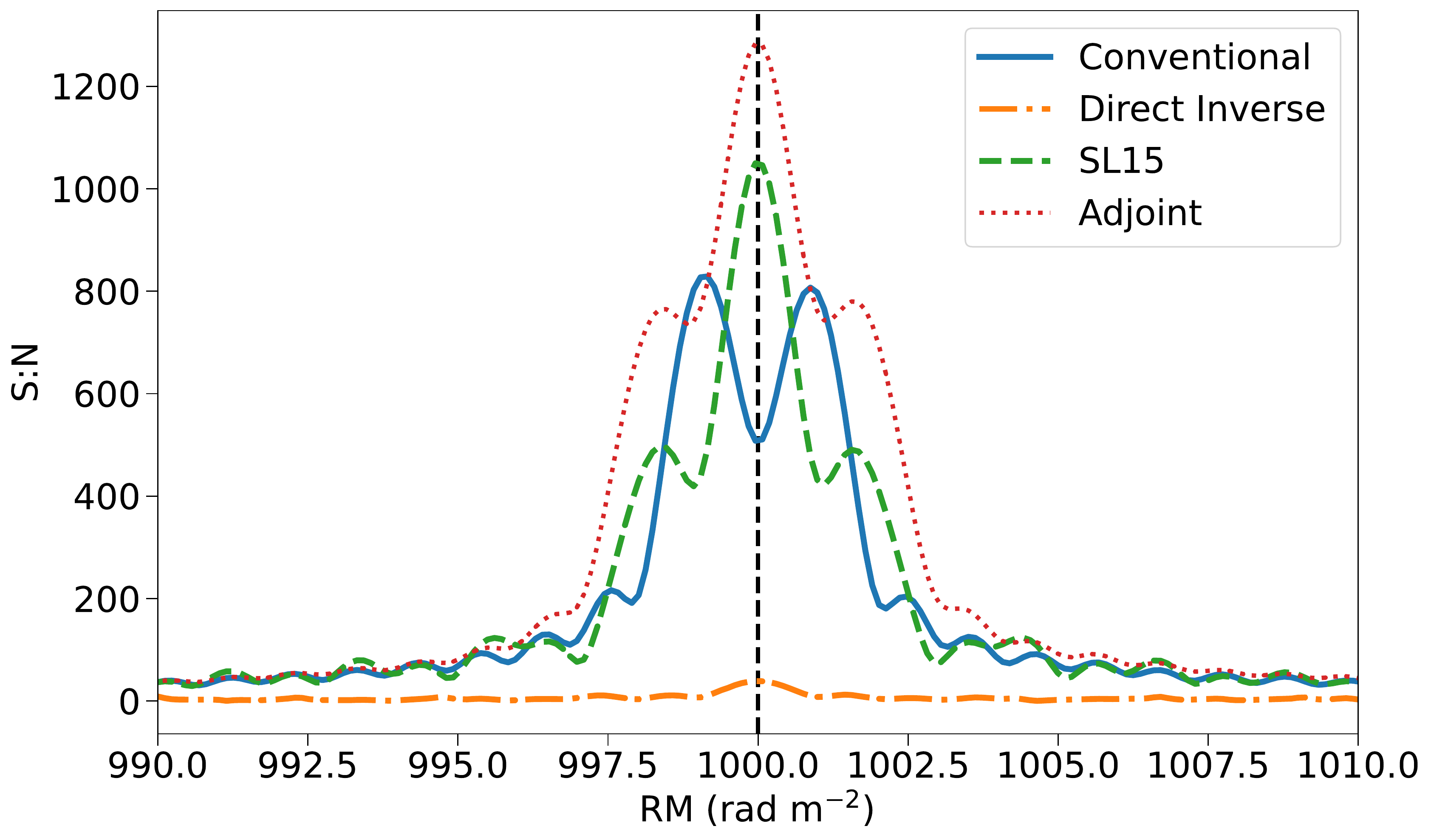}
    \caption{The FDFs from Fig~\ref{fig:Faraday Disperson Functions}, normalized by their respective theoretical noise curves so as to be a signal-to-noise ratio (S:N) and shown on a common axis. The direct inverse transform greatly amplifies the noise, causing it to have low S:N. The adjoint transform produces the highest S:N at the peak. The true RM (the RM of the simulated source) is marked by a black dashed line.}
    \label{fig:snr_comparison}
\end{figure}

Since the noise is a function of Faraday depth for the direct inverse and adjoint transforms, it is not possible to directly assess the significance of a peak in an FDF such as is normally done with conventional RM-synthesis. Since the noise behaviour of the transforms follows the theoretical predictions, it is possible to normalize the FDFs by the noise curves to generate modified FDFs with (nominally) unit noise, effectively giving the S:N across the entire FDF and making it straightforward to determine the significance of a peak. Fig.~\ref{fig:snr_comparison} shows the FDFs from Fig.~\ref{fig:Faraday Disperson Functions}, zoomed in on the true RM and normalized by the theoretical noise functions. As expected based on the previous results, the direct inverse transform's greatly amplified noise causes its S:N to be very poor, the \citetalias{Schnitzeler2015} transform improves on the conventional transform, and the adjoint transform produces the best S:N. This is a general result: for all RMs (other than zero, where all transforms become the same), the adjoint transform will produce the highest S:N peak.

\section{Algorithm development} \label{sec:algorithm}
Based on the results of the previous section, the modified transforms are effective at producing a peak at the location of the true RM and correctly predicting the amount of bandwidth depolarization, allowing the correct polarized intensity to be recovered. However, it is not guaranteed that the peak at the location of the true RM is the highest amplitude peak, in either the FDF or the sensitivity-corrected FDF. Thus, a modified method of identifying which peak(s) to analyze is required. Instead of using the highest amplitude peak in the FDF to determine the RM peak (or assigning a threshold and analyzing all peaks above that threshold), the highest peak (or a threshold) in the S:N can be used. This method accounts for RM-dependent noise in the FDF, and is far less likely to register a false peak as the most significant. This method also has the advantage of being very easy to assign significance thresholds for, as astronomers typically define such thresholds in units of S:N.

Once the strongest peak in S:N is located, the FDF (or only a segment around the peak) can be normalized by the sensitivity curve, and the peak can be characterized as normal to determine the RM, polarized intensity, and polarization angle of the source.

We selected the adjoint transform as the best candidate for use in this algorithm, as it produces the highest S:N and reliably follows the theoretical sensitivity and noise curves. The adjoint transform should give the best chance of detecting a strongly bandwidth-depolarized source while allowing detected sources to be measured accurately. We developed a modified version of the \textsc{RM-Tools} \texttt{rmsynth1d} tool that uses this algorithm.

\subsection{RMSF estimation}
As mentioned previously, the RMSF is no longer well defined for the new transforms. The RMSF is needed for three reasons: visual assessments of Faraday complexity, estimating the uncertainty in RM and other quantities, and deconvolution algorithms such as RM-Clean \citep{Heald2009}. The first and last of these are outside the scope of this work. But the estimation of uncertainties is important even for the case of Faraday-simple sources, so an equivalent to the RMSF is needed.

It is straightforward to define a modified RMSF, $\widetilde{R}(\phi,\mathrm{RM})$, by beginning with the premise that this modified, RMSF function shows the FDF of an ideal Faraday-simple source with unit amplitude and zero polarization angle (effectively, $\widetilde{P}_0 = 1$). In conventional RM synthesis, the RMSF is same as this, but explicitly for an RM of zero ($\mathrm{RMSF}(\phi) = \widetilde{R}(\phi,0)$). It can be straightforwardly derived using Fourier properties that the modified version for the point-response, simulating the ideal source of arbitrary RM, is simply a shifted version of the RMSF: $\widetilde{R}(\phi,\mathrm{RM}) = \mathrm{RMSF}(\phi-\mathrm{RM})$. This property is exploited by the RM-Clean algorithm to calculate the point-response of an arbitrary clean component.

For the modified transforms we cannot use the Fourier properties to make this simplification, so we can only define the modified RMSF in terms of the modified transform itself. In the case of the adjoint transform, this can be done by setting $\widetilde{P}_0 = 1$ in Eq.~\ref{eq:adjointFDF}:
\begin{equation}\label{eq:adjointRMSF}
    \widetilde{R}_\mathrm{adj}(\phi) = \left(\frac{1}{\sum_i w_i}\right) \sum_i w_i R_i(\nu,\mathrm{RM}) \, R_i^*(\nu,\mathrm{\phi}).    
\end{equation}
For any given RM, the modified RMSF can be computed using this equation, and then can be used similarly to a normal RMSF with the exceptions that the peak occurs at $\phi = \mathrm{RM}$ instead of at zero and the peak amplitude will generally not be one. The width of the modified RMSF, which is used in the error analysis, can be computed in the usual way, and in principle the modified RMSF could be used for deconvolution without requiring the shift step used in conventional RM-Clean In other words, a local RMSF must be computed rather than using a shifted version of the conventional RMSF.

The RM-dependent RMSF does introduce some additional complications. An FDF typically has the same units as the supplied channel data, but with the addition of an RMSF$^{-1}$ factor (analogous to how radio intensity images typically have units of Jy/beam). For Faraday-simple polarization features, the polarized intensity can be read from the peak of the FDF for both conventional RM synthesis and the adjoint method. But attempting to integrate over an FDF, e.g. to determine the integrated polarized intensity of a Faraday-complex feature, becomes more difficult as this must account for the variable shape of the RMSF with changing RM (although we note that this is probably not a strong effect so long as the integration range in RM is narrow compared to the magnitude of RM). We also note that the adjoint RMSF can have broader and/or taller sidelobes as a result of the effective weighting, which can further complicate any studies of Faraday complexity in strongly-bandwidth depolarized sources. A detailed analysis of these problems is beyond the scope of this work, but these problems provide additional motivation for choosing observational configurations that will minimize bandwidth depolarization when planning studies of Faraday complexity.

\subsection{Error Analysis}\label{sec:error}
\textsc{RM-Tools} generates uncertainties for the polarized properties it reports, using a set of standard equations which depend predominantly on the noise in the FDF and the width of the RMSF. We assumed that the same equations still apply to the modified algorithm, but we had to rework the uncertainty estimation slightly because the RMSF and noise are now RM-dependent. For the RMSF width, the existing method of performing a Gaussian fit to the main lobe of the RMSF was still effective, except for a minor code change to allow it to change the location of the RMSF peak. The RMSF width's dependence on RM has the consequence that the theoretical width calculated by \textsc{RM-Tools} is not accurate at RMs far from zero -- the width must be fit in order for the corresponding uncertainties to be accurate. 

The theoretical noise in the (sensitivity-corrected) FDF was modified to use Eq.~\ref{eq:adjointnoise}, calculated at the Faraday depth of the peak and normalized by the sensitivity (Eq.~\ref{eq:adjointsensitivity}) at the same Faraday depth. Unlike conventional RM-synthesis, the option to estimate the actual noise in the FDF using apparently signal-free regions away from the peak is not practical in this modified algorithm due to the Faraday-depth dependence in the noise.

To test the accuracy of the uncertainties reported by the modified algorithm, we constructed a series of simulations and analyzed the residual differences between those simulation inputs and the outputs of our algorithm; the details of these simulations are described in Appendix~\ref{app:err}. We found that across the broad range of S:N and bandwidth depolarization values that we tested the reported uncertainties were fairly accurate: the polarized intensity errors were consistent with the reported uncertainties; the normalized RM errors were 20\% smaller, indicating that the uncertainties were slightly too large; the de-rotated polarization angle errors were 15\% smaller than expected, indicating that the uncertainties were underestimated. The RM error discrepancy was found to be caused by the uncertainties at large RMs (where the bandwidth depolarization is stronger than 80\%) being significantly overestimated; the RM uncertainties are correct where the bandwidth depolarization is less extreme than this.

\subsection{Software implementation}
We have created a modified version of \textsc{RM-Tools}' \texttt{rmsynth1d} script to implement the bandwidth depolarization correction algorithm described above. This is not intended to serve as a general replacement to \texttt{rmsynth1d}, as most data is generally expected to not be significantly affected by bandwidth depolarization and the higher computational cost and loss of Fourier formalism are likely not worth the change in most cases.

Instead, we chose to offer this modified script as an alternative option available to users who wish to search for possible bandwidth depolarized sources in their data. We have called our new script \texttt{rmtools\_bwdepol}, and had it incorporated into \textsc{RM-Tools} as of version 1.3.

The \texttt{rmtools\_bwdepol} can take an input file with the same format as \texttt{rmsynth1d}, in which case it assumes the channel widths based on the difference between the frequencies of the first two channels (all channels are assumed to have the same width). Alternatively, the input file can have an extra column giving the channel width in Hz. The tool produces the same outputs as \texttt{rmsynth1d}, except that the plotted RMSF is the modified RMSF local to the identified peak, plus an additional plot is given showing the theoretical sensitivity and noise as functions of Faraday depth. The reported uncertainties have been changed as described in the previous sub-section, and a warning has been added for cases where the bandwidth depolarization is so strong that the reported uncertainties may be unreliable. Full documentation on the use of this script, and \texttt{rmsynth1d}, can be found within the script, on the \textsc{RM-Tools} wiki,\footnote{https://github.com/CIRADA-Tools/RM-Tools/wiki} and in the forthcoming \textsc{RM-Tools} description paper (Van Eck et al., in prep). The compute time of this tool, compared with \texttt{rmsynth1d}, are about 2-5 times longer for the same data and run parameters, as a result of the additional mathematical complexity of the transform. The typical run time for \texttt{rmtools\_bwdepol} is likely to be somewhat longer than that, as the typical use case will be to compute larger Faraday depth ranges than would typically be used for \texttt{rmsynth1d}.

We have also created a second script, \texttt{rmtools\_bwpredict}, which shows users the theoretical sensitivity and noise curves for a given channel configuration. This is intended to provide an easy way for users to see if bandwidth depolarization is a potential problem for either a given dataset or planned future observations, and as a more accurate calculation than the empirical relation in Eq.~\ref{eq:C}. Users can use the results to decide whether it is worth using the \texttt{rmtools\_bwdepol} tool to search for extreme RMs and/or determine depolarization-corrected polarization properties from sources already known to have extreme RMs.

The script takes as input a text file containing the list of channel frequencies and, optionally, the channel widths. If the widths are not included, they are assumed to be the same for all channels and are taken from the separation of the first two channels.

%\begin{figure*}
%	% Allowable file formats are eps or ps if compiling using latex
%	% or pdf, png, jpg if compiling using pdflatex
%	\includegraphics[angle=90,width=\textwidth,height=0.9%\textheight,keepaspectratio]{f6.pdf}
%    \caption{ {\it Panel C:} Adjoint transform FDF ({\it bottom}) normalized by sensitivity and modified RMSF ({\it top}), with the highest SNR peak marked by the pink dashed line. {\it Panel D:} as panel C, but zoomed into the peak.
%    }
%    \label{fig:output_example}
%\end{figure*}
\begin{figure*}%[h!]
    \centering
\begin{subfigure}{1.0\textwidth}
    \centering
    \includegraphics[width=\textwidth,height=0.4\textheight,keepaspectratio]{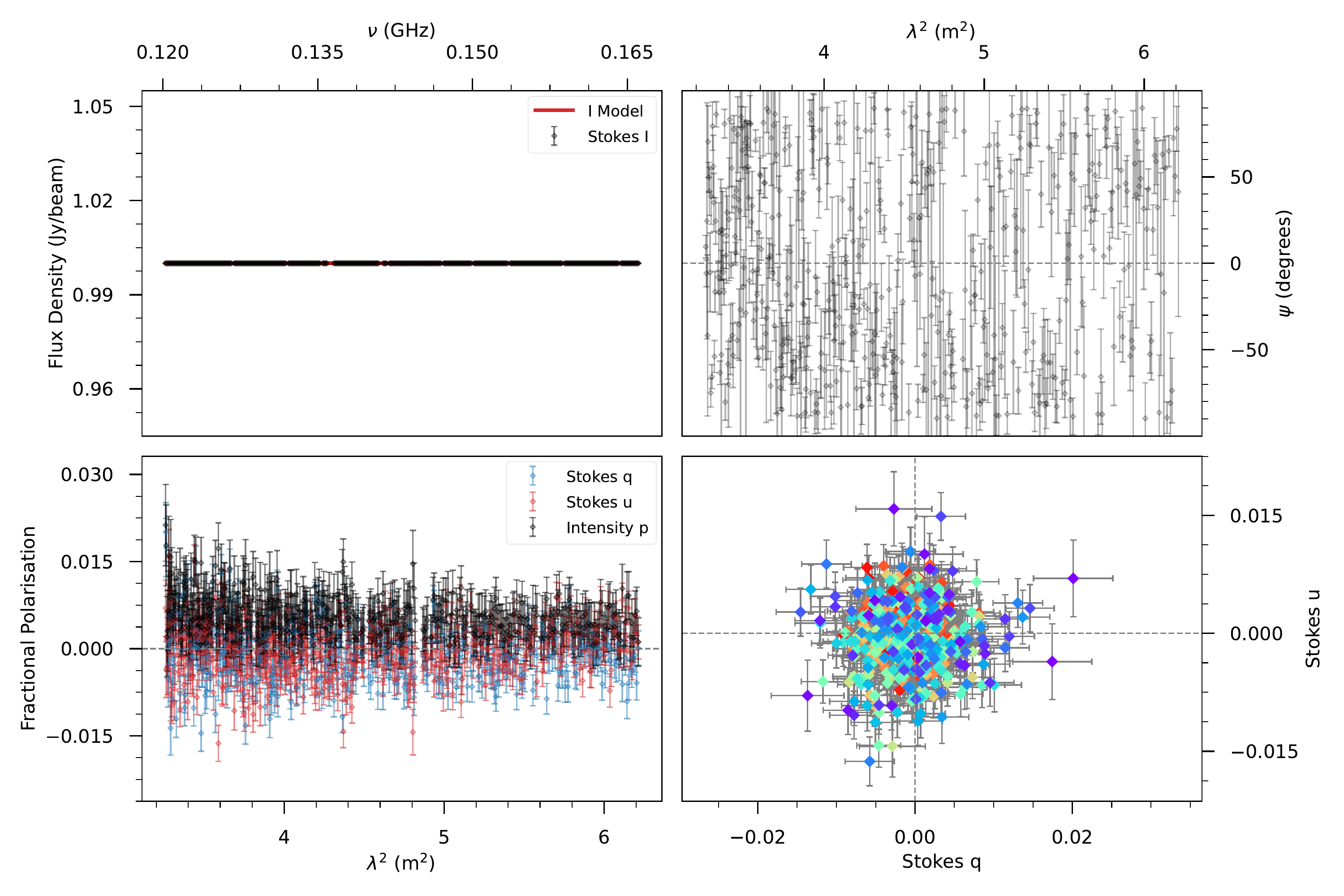}
    \caption{The frequency domain polarization properties: Stokes I spectrum and model ({\it top left}), fractional Stokes $Q$, $U$, and polarized intensity ({\it bottom left}), polarization angle ({\it top right}), and $q-u$ plane ({\it bottom right}). These are unchanged from the default RM-Tools behaviour.}
    %\label{fig:first}
\end{subfigure}

\begin{subfigure}{\textwidth}
    \centering
    \includegraphics[width=\textwidth,height=0.4\textheight,keepaspectratio]{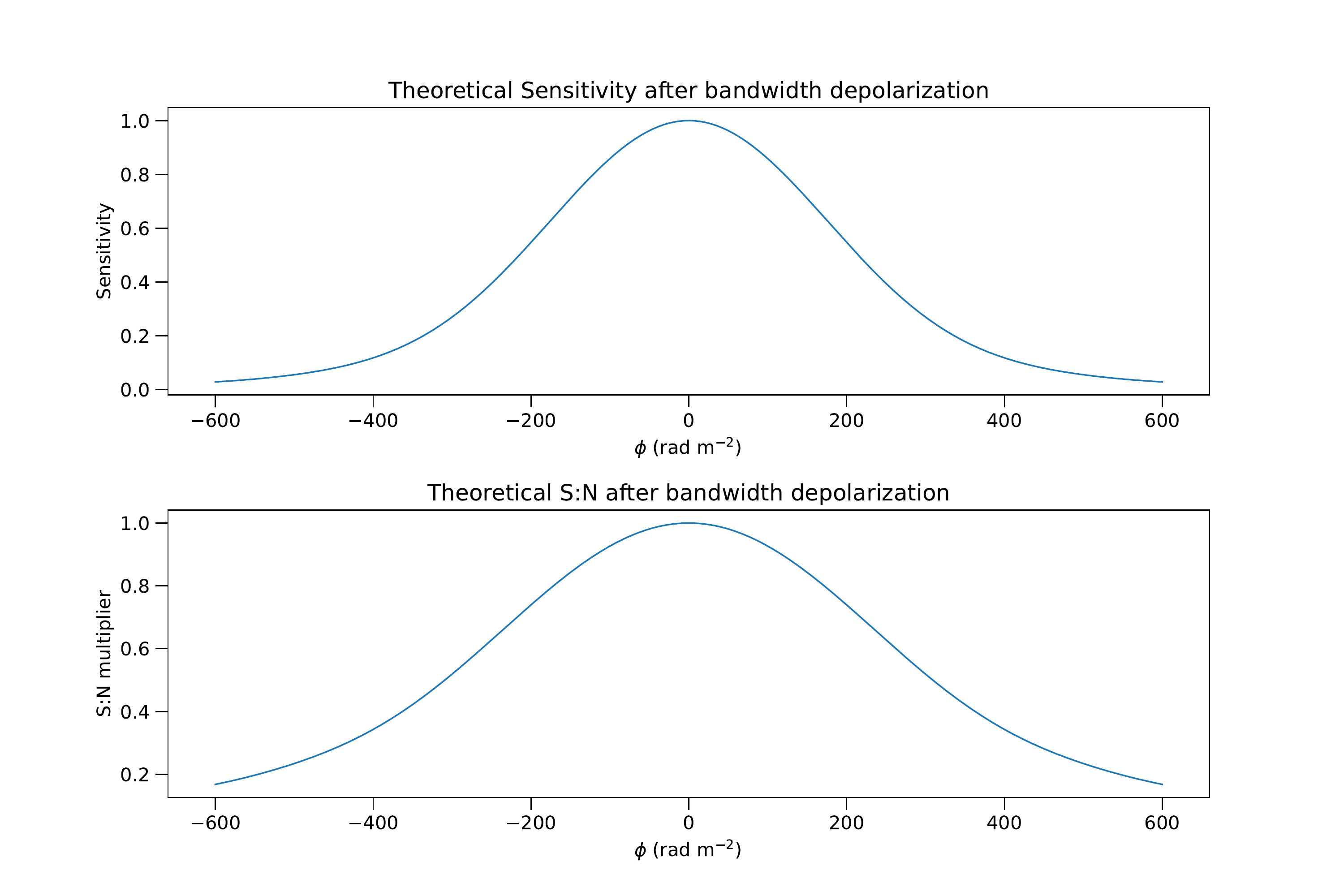}
    \caption{Theoretical sensitivity ({\it top}) and noise ({\it bottom}) curves for the adjoint transform.}
    %\label{fig:third}
\end{subfigure}
\caption{Example outputs of the rmtools\_bwdepol tool, showing one of the LoTSS sources.}
\label{fig:output_example}
\end{figure*}
%%%%%%%% Continue figures %%%%%%%%
\begin{figure*}
\ContinuedFloat
    
\begin{subfigure}{\textwidth}
    \centering
    \includegraphics[width=\textwidth,height=0.4\textheight,keepaspectratio]{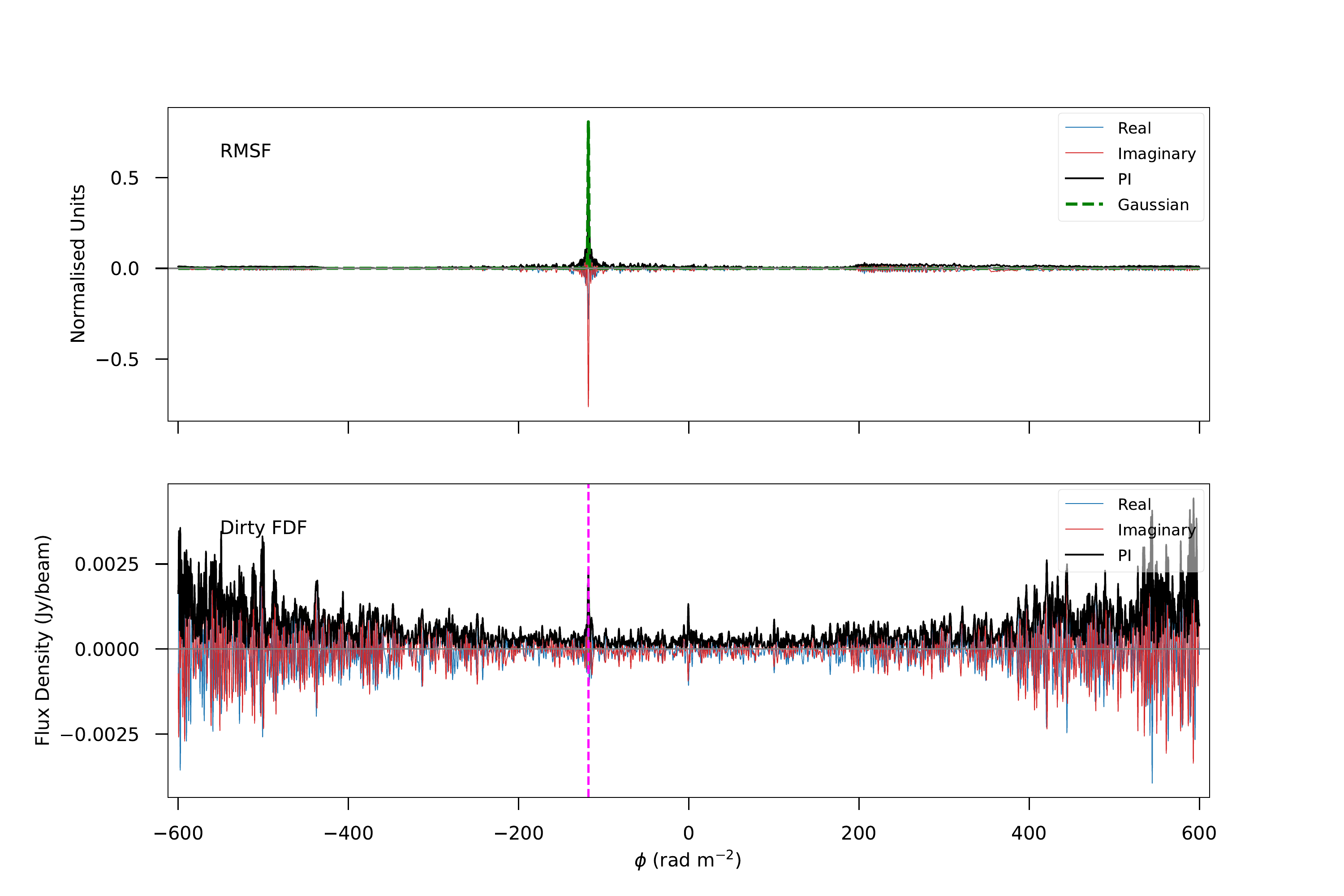}
    \caption{Adjoint transform FDF ({\it bottom}) normalized by sensitivity and modified RMSF ({\it top}), with the highest S:N peak in the FDF marked by the pink dashed line.}
    %\label{fig:third}
\end{subfigure}
\begin{subfigure}{\textwidth}
    \centering
    \includegraphics[width=\textwidth,height=0.4\textheight,keepaspectratio]{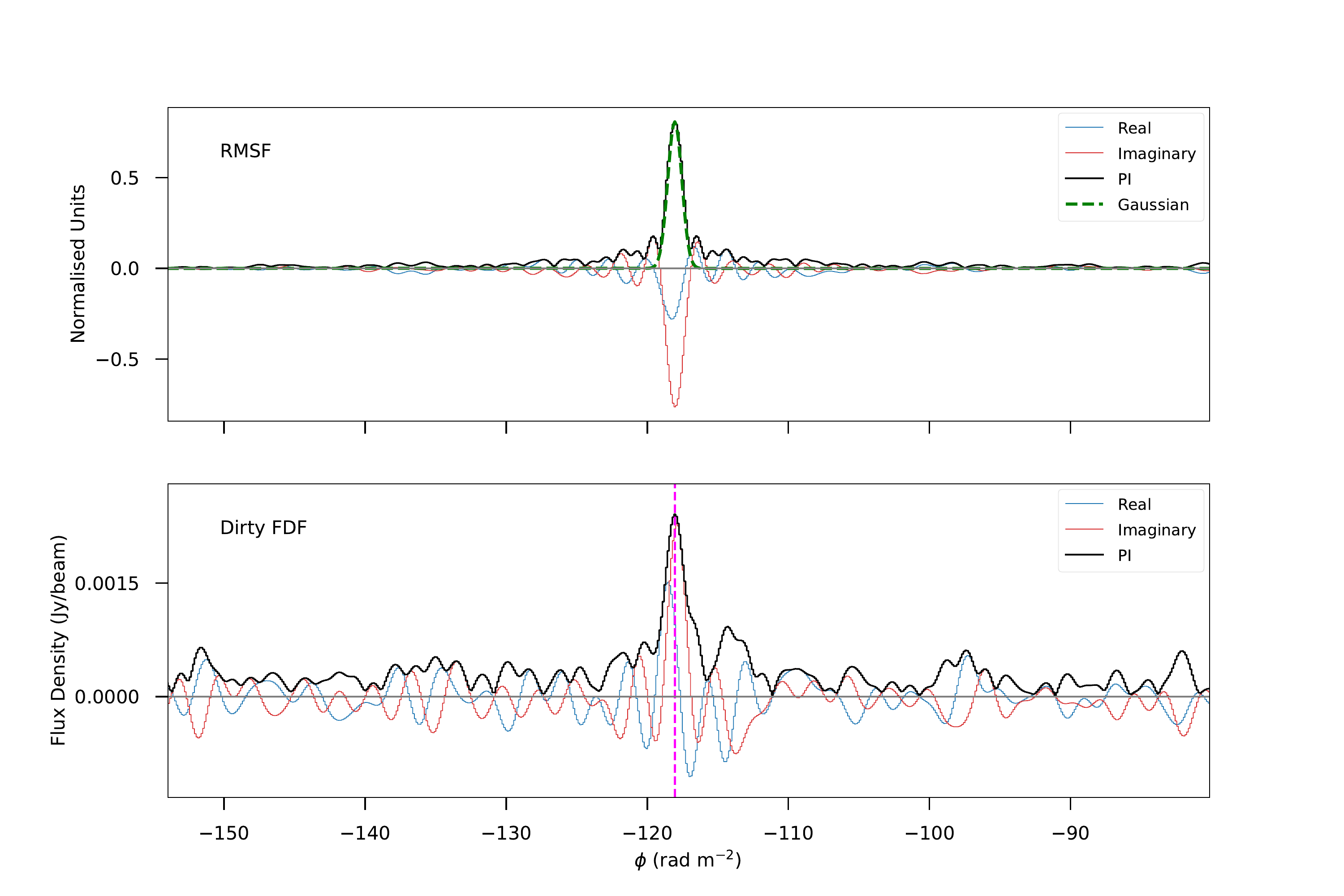}
    \caption{As panel (c), but zoomed into the peak.}
    %\label{fig:third}
\end{subfigure}        
\caption{ continued. Example outputs of the rmtools\_bwdepol tool, showing one of the LoTSS sources.}
\label{fig:output_example}
\end{figure*}

\section{Tests on real data}\label{sec:tests}
To verify the operation of our modified script, as well as test the performance of the modified transforms on actual data, we obtained data from the LoTSS second data release \citep{Shimwell2022}.
For the LoTSS frequency configuration (120 -- 168 MHz, 97.6 kHz channels), bandwidth depolarization starts to become significant for RMs above approximately 100 \rmu. From the LoTSS DR2 polarization catalog \citep{O'Sullivan2023}
19 sources with $|\mathrm{RM}| > 100$ \rmu\ were identified, and their polarized spectra obtained. We processed these sources using both conventional RM-synthesis (as implemented in \textsc{RM-Tools}), and our \texttt{rmtools\_bwdpol} script. Fig.~\ref{fig:output_example} shows an example of one of these sources as shown by \texttt{rmtools\_bwdpol}.

For all sources except one, we found that both methods recovered the same RM, and approximately similar values for the fractional polarization and polarized intensity. The single exception was found to be a source where the true polarized component was only very slightly stronger than the instrumental leakage peak at RM $\sim$ 0 \rmu. In the adjoint method the comparatively reduced sensitivity at large RM values caused the true peak to appear as slightly lower SNR compared to the leakage peak, resulting in the script identifying the leakage peak as the most significant. When the leakage peak was blanked from the FDF, the peak-finding algorithm correctly identified the same peak as for the conventional FDF.

\begin{figure}
	\includegraphics[width=\columnwidth]{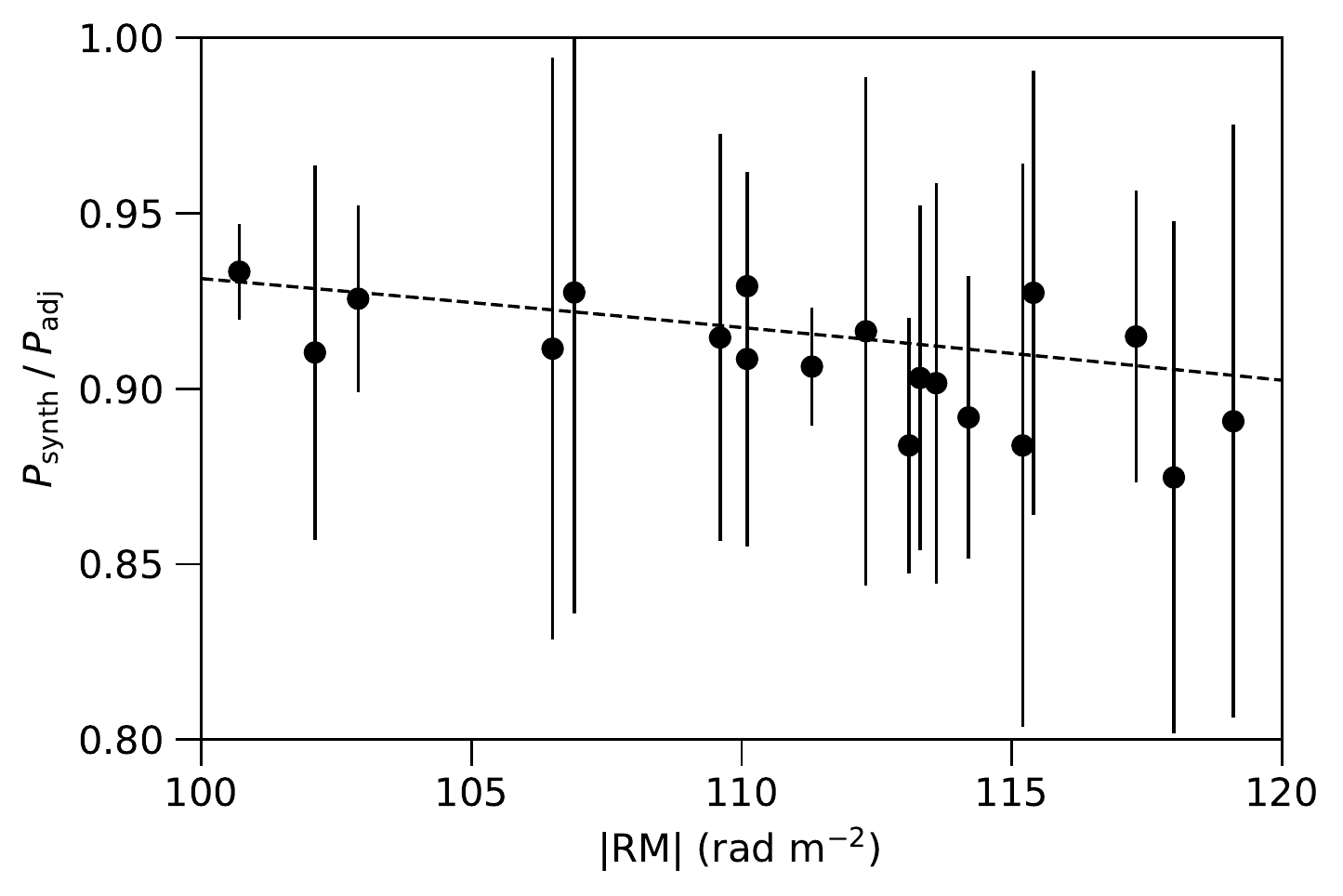}
    \caption{The ratio of measured polarized intensities using the conventional RM synthesis method and the adjoint method, as a function of measured RM. The error bars in the ratio were computed assuming the errors were independent between methods, which likely overestimates the true errors; the errors in RM are smaller than the symbol size. The dashed line is the theoretical upper limit for the bandwidth depolarization predicted by Eq.~\ref{eq:Q}.}
    \label{fig:LOFARdepol}
\end{figure}

To confirm that the modified method was correcting the polarized intensity (and fractional polarization) as expected, we checked that the polarized intensities measured by the conventional method were lower than those from the adjoint method, and by the expected amount; this is shown in Fig.~\ref{fig:LOFARdepol}. 
The data appear to follow the theoretical prediction (which is an upper limit, but for this degree of depolarization one that should be fairly accurate) quite well, and show that for these sources the bandwidth depolarization is predicted to be strong enough to cause the measured polarized intensities to be significantly underestimated, at the level of greater than 1$\sigma$. Since we do not know the true polarized intensity for these sources, we cannot confirm that the adjoint method recovers the correct values, but we can confirm that the method does produce values consistent with the theoretical expectations.

\section{Discussion and conclusions}\label{sec:conclusions}
Bandwidth depolarization occurs when we observe Faraday-rotated polarized emission using frequency channels with finite bandwidth. Most observations are configured to minimize this effect so that it can be safely ignored, but in some cases the effects can be significant enough to affect measurements of polarized intensity and RM, and may reduce the signal-to-noise ratio in polarization below the threshold for detection. This can lead to observational bias by both causing observations to miss polarized sources with high-magnitude RMs, or by causing those sources to appear significantly less polarized than they actually are. We predict that bandwidth depolarization may be significant for the LoTSS survey near the Galactic plane ($|\mathrm{RM}| \gtrsim$ 100 \rmu), and for the VLASS coarse-channelization data in regions of very strong Faraday rotation ($|\mathrm{RM}| \gtrsim$ 1000 \rmu). It's also quite likely that future polarization surveys with the SKA will encounter problems with bandwidth depolarization. Thus is it beneficial to develop an algorithm that can reduce these biases as much as possible for observations where this may be a problem.

We have found a simple equation that can give observers a method to estimate the strength of bandwidth depolarization for a given channel configuration. This equation can be used to work out the range of RMs in which bandwidth depolarization is below a chosen threshold, or could be reversed to determine the approximate strength of depolarization at a selected RM, so long as it remains within the weak-depolarization regime (less than 50\% loss of polarized signal). More accurate values, particularly in the case of strong depolarization, require numerical calculations with the more accurate model.

We have laid out a mathematical model for how bandwidth depolarization operates in conventional RM synthesis algorithms, including an upper limit for the ratio of measured polarized intensity to true polarized intensity due to bandwidth depolarization (Eq.~\ref{eq:Q}). An exact solution for this ratio does not appear to be possible, therefore it is not possible to derive a correction for the depolarization in conventional RM synthesis. In addition, we have confirmed previous results that strong bandwidth depolarization can cause the measured RM to be incorrect due to false Faraday complexity manifesting in the Faraday depth function.

We explored three modified versions of RM synthesis, replacing the Fourier kernel with alternative functions that account for bandwidth depolarization in various ways. We derived, and then tested through simulations, the behaviour of these modified transforms to determine whether they could be used to produce better results than conventional RM synthesis. We have verified that the behaviour is consistent with the theory for several of the channel configurations in present use, and expect the theoretical predictions to hold for a variety of possible configurations.

The adjoint transform produces the highest signal-to-noise of the four transforms we tested, and it can recover the true polarized intensity given the well-behaved sensitivity curve. We therefore selected this transform for our modified RM synthesis algorithm. This algorithm functions similarly to conventional RM synthesis, but uses the adjoint transform and applies the predicted noise function to normalize the FDF into a signal-to-noise ratio. This can then be used to identify statistically significant peaks indicating a polarized source. If such a peak is found, the predicted polarized intensity sensitivity function is used to renormalize the FDF into units of depolarization-corrected polarized intensity, and this FDF is used to characterize the polarization properties as normal. 

We have implemented this algorithm within the \textsc{RM-Tools} package as an alternative to the conventional RM-synthesis tool. We have also developed a tool for predicting the strength of bandwidth depolarization effects as a function of RM for arbitrary user-defined frequency configurations, which can be used to assist in determining whether the use of this modified algorithm is merited for any given data set. These tools are now available to the community as part of the \textsc{RM-Tools} package under the names \texttt{rmtools\_bwdpol} and \texttt{rmtools\_bwpredict}.

Using data from the LoTSS survey, we tested the results of this modified algorithm in comparison to conventional RM synthesis. For large RMs ($|\mathrm{RM}| > 100$ \rmu) we predict that the depolarization is significant (>5\% loss of polarized intensity) for sources in these observation. Our algorithm recovers the same RMs as conventional RM synthesis and polarized intensities that are larger by the expected amount, which indicates that the correction is working as intended.

We propose that this method will be useful for those data sets where bandwidth depolarization will be significant. We have derived all of these results for idealized Faraday-simple sources, so it may be of interest in the future to extend the theory and methods to include Faraday-complex sources and deconvolution methods such as RM-Clean.

\section*{Acknowledgements}

This research made extensive use of \textsc{Astropy}, a community-developed core Python package for Astronomy \citep{Astropy,Astropy2}; \textsc{SciPy} \citep{Scipy}; \textsc{NumPy} \citep{Numpy}; \textsc{matplotlib} \citep{Matplotlib}; and \textsc{RM-Tools} \citep{Purcell2020}. The authors thank Shane O'Sullivan for his assistance in providing the LoTSS data prior to its public release, and the referee for their careful review and thoughtful comments.

%%%%%%%%%%%%%%%%%%%%%%%%%%%%%%%%%%%%%%%%%%%%%%%%%%
\section*{Data Availability}
This work has made use of publicly-available data from the LOFAR Two-metre Sky Survey (LoTSS) data release 2 \citep{Shimwell2022}. These data are available online at \url{doi.org/0.25606/SURF.LoTSS-DR2}. The polarization catalog and spectra that were used are described in \citet{O'Sullivan2023} and are available online at \url{https://lofar-mksp.org/data/}.



\bibliographystyle{mnras}
\bibliography{References} % if your bibtex file is called example.bib

%\begin{thebibliography}{99}
%\end{thebibliography}

%%%%%%%%%%%%%%%%%%%%%%%%%%%%%%%%%%%%%%%%%%%%%%%%%%

%%%%%%%%%%%%%%%%% APPENDICES %%%%%%%%%%%%%%%%%%%%%

\appendix
\section{Error analysis simulations} \label{app:err}
As described in Sec.~\ref{sec:error}, the uncertainty estimates were modified from the \textsc{RM-Tools} defaults to account for the changes made to the RM-synthesis by the adjoint method. The RMSF width, which is used in the RM uncertainty calculation, was changed to always use the measured value, as the theoretical value significantly underestimates the actual width at large RMs. The uncertainty in the polarized intensity and derotated polarization angle were modified to account for both the RM-dependent noise and the sensitivity correction.

To test the accuracy of these revised uncertainties, we constructed a series of simulations. We used the same LOFAR-like channel configuration described in Sec.~\ref{sec:comparison}. For each simulation iteration, random values for the polarized intensity, RM, and initial polarization angle were assigned: polarized intensity was drawn uniformly between zero and 3 arbitary flux units; RM was drawn uniformly in the range [-500, 500) \rmu, which corresponded to a maximum possible depolarization of 97\% (3\% remaining polarized intensity); and the initial polarization angle was drawn uniformly in the range [-90,90) degrees. From these parameters, the Stokes $Q$ and $U$ values of a ideal Faraday-thin source were computed. To each $Q$ and $U$ value, random noise was drawn from a Gaussian distribution with unit variance.

The corresponding range of possible band-averaged S:N values, neglecting bandwidth depolarization, was from zero to 67. In practice, bandwidth depolarization reduced the S:N, producing a significant number of iterations with very low S:N. Iterations with a measured S:N of the final peak of less than 8 were rejected as non-detections. New iterations were generated until the total number of detected simulated sources was 10 000.

\begin{figure}
	\includegraphics[width=\columnwidth]{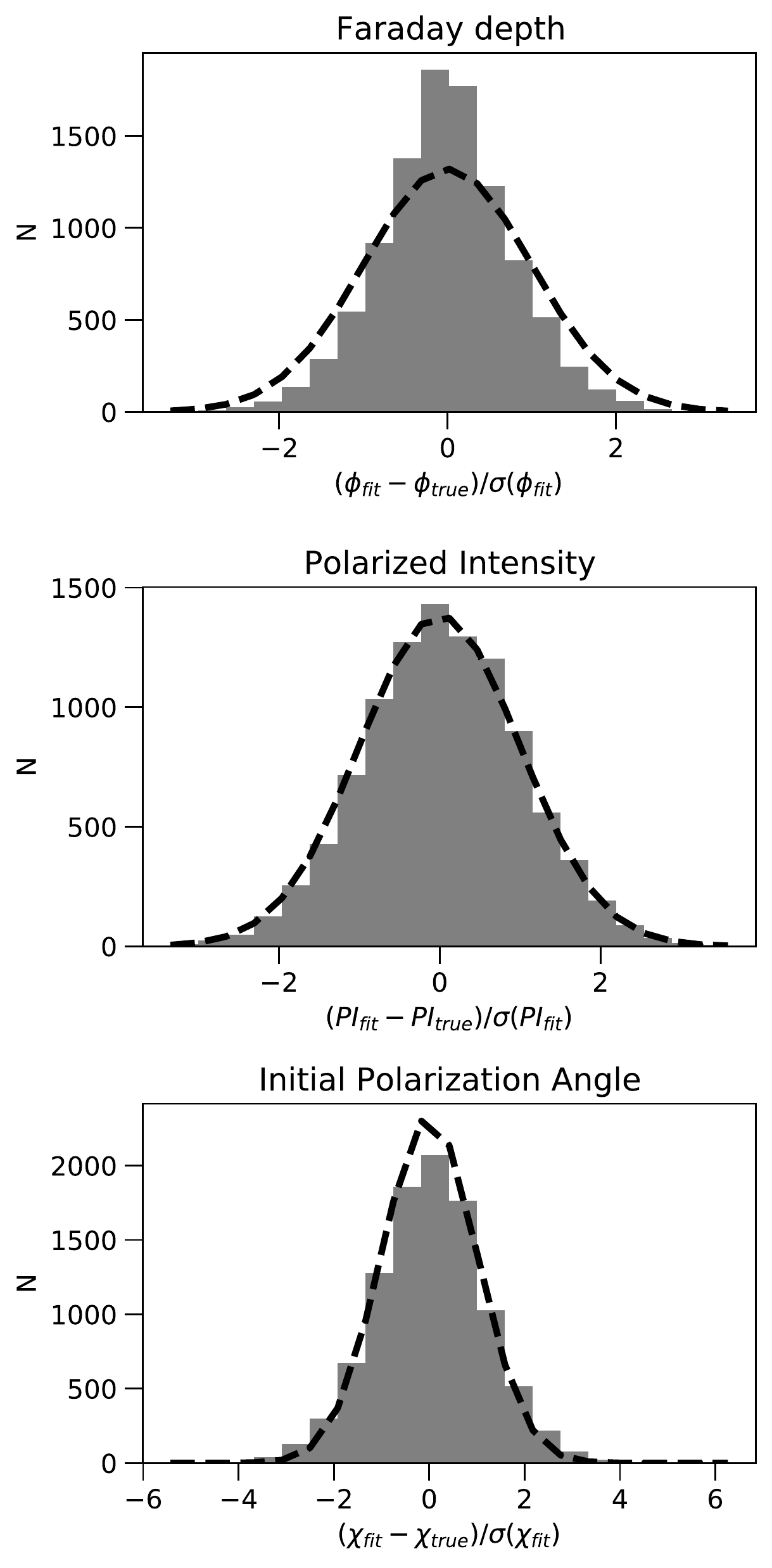}
    \caption{Histograms of uncertainty-normalized residuals for RM (top), polarized intensity (middle) and polarization angle (bottom), generated from 10~000 simulations of polarized sources in the LoTSS frequency coverage. The residual distribution of RM is narrower than expected, indicating the reported uncertainties are too large; the polarized intensity distribution matches expectations, indicating good uncertainty estimates; the derotated polarization angle distribution is slightly too wide, indicating slightly too-small reported uncertainties.
      }
    \label{fig:error_sims}
\end{figure}

For all of the iterations, we computed the uncertainty-normalized residuals in polarized intensity, RM and de-rotated polarization angle, by calculating the difference between the adjoint-method output value and the simulation ground truth and then normalizing by the reported uncertainty. The resulting distributions are shown in Figure~\ref{fig:error_sims}. The ideal outcome is that the normalized residual have a Gaussian distribution of unit variance, indicating that the errors are well behaved and are well characterized by the reported uncertainties. Our simulations show that the errors are Gaussian distributed, but the widths deviate slightly from the ideal. The RM residual distribution is narrower than expected (by 20\%), indicating that the uncertainties are slightly overestimated. The polarized intensity is very close to ideal, indicating that the uncertainties are within a few percent of correct. The polarization angle distribution is slightly broader than expected (10\%), indicating that the uncertainties are slightly underestimated.

\begin{figure}
	\includegraphics[width=\columnwidth]{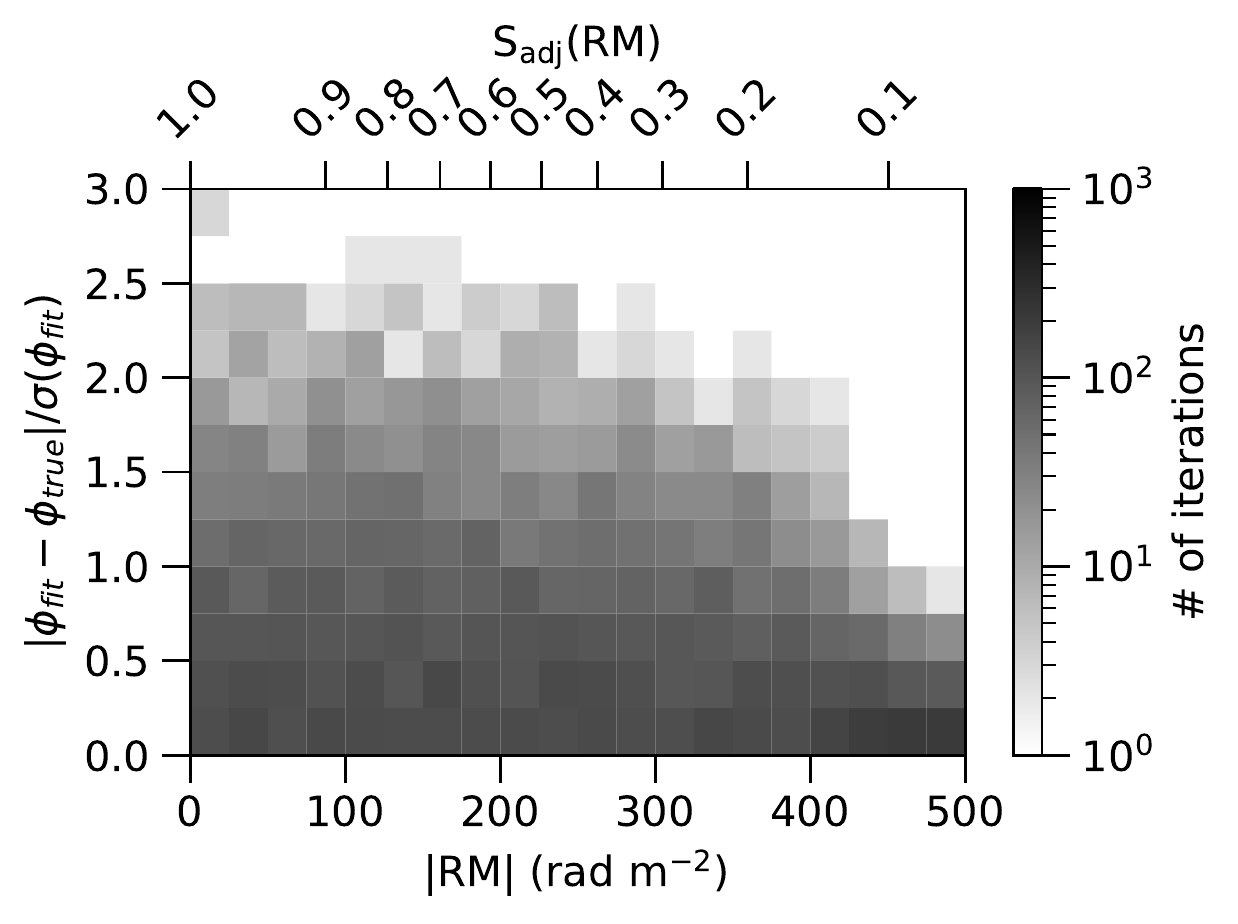}
    \caption{Distribution of normalized residuals as a function of RM (or sensitivity). A strong decrease in the residual variance can be seen below a sensitivity threshold of 20\%, indicating the uncertainties in that regime are significantly overestimated.
      }
    \label{fig:error_distribution}
\end{figure}

We investigated the behaviour of the RM errors and found that the normalized residuals decreased significantly for very large RMs, where the depolarization becomes very strong (more than 80\% loss of polarized signal). Figure~\ref{fig:error_distribution} shows this effect, as well as how the normalized residuals are approximately constant in the less-depolarized regime. We investigated if this effect scales with the sensitivity (and not RM) by repeating the simulations using a higher frequency channel configuration (800 - 1088 MHz, with 1 MHz channels). We confirmed that when the sensitivity dropped below 0.2 (80\% depolarization) the uncertainties became significantly underestimated. We conclude that the RM uncertainties are accurate to within a few percent for the regime where the sensitivity is greater than 20\%, and the RM uncertainties are overestimated by potentially a factor of a few when the sensitivity is below this value. We repeated this analysis on the polarized intensity and polarization angle, and found that their residuals show no trends with RM or depolarization fraction.

%%%%%%%%%%%%%%%%%%%%%%%%%%%%%%%%%%%%%%%%%%%%%%%%%%

% Don't change these lines
\bsp	% typesetting comment
\label{lastpage}
\end{document}